\documentclass[journal, 10pt,twocolumn]{IEEEtran}
\usepackage{cite}
\usepackage{amsmath, amssymb, amsfonts, amsthm}
\usepackage{algorithm}
\usepackage[noend]{algorithmic}
\usepackage{graphicx}
\usepackage{textcomp}
\usepackage{xcolor}
\usepackage{color}
\usepackage{bm} 
\usepackage{subfigure} 
\usepackage{booktabs}
\usepackage{autobreak}
\usepackage{multirow}
\usepackage{booktabs}
\usepackage{authblk}
\usepackage{mathrsfs}
\usepackage{array}
\usepackage{makecell} 
\usepackage{float}
\usepackage{soul}
\usepackage{stmaryrd}
\usepackage{dsfont}
\usepackage{bbm}
\usepackage[hidelinks]{hyperref}
\usepackage{cleveref}

\definecolor{blue}{RGB}{0,0,0}

\newtheorem{theorem}{Theorem}

\begin{document}

\title{Federated Inference with Reliable Uncertainty Quantification over Wireless Channels via Conformal Prediction}

\author{
Meiyi Zhu, Matteo Zecchin \IEEEmembership{Student Member, IEEE}, Sangwoo Park \IEEEmembership{Member, IEEE}, Caili Guo \IEEEmembership{Senior Member, IEEE}, Chunyan Feng \IEEEmembership{Senior Member, IEEE}, Osvaldo Simeone \IEEEmembership{Fellow, IEEE}

\vspace{-0.5cm}

\thanks{
The work of M. Zhu, C. Guo and C. Feng was supported by the National Natural Science Foundation of China (62371070), by the Fundamental Research Funds for the Central Universities (2021XD-A01-1), and by the Beijing Natural Science Foundation (L222043). The work of M. Zhu was also supported by the BUPT Excellent Ph.D. Students Foundation (CX2023150).
The work of M. Zecchin and O. Simeone was supported by the European Union's Horizon Europe project CENTRIC (101096379). The work of O. Simeone was also supported by an Open Fellowship of the EPSRC (EP/W024101/1), by the EPSRC project (EP/X011852/1), and by Project REASON, a UK Government funded project under the Future Open Networks Research Challenge (FONRC) sponsored by the Department of Science Innovation and Technology (DSIT).

Meiyi Zhu, Caili Guo and Chunyan Feng are with the Beijing Key Laboratory of Network System Architecture and Convergence, School of Information and Communication Engineering, Beijing University of Posts and Telecommunications, Beijing 100876, China (e-mail: lia@bupt.edu.cn; guocaili@bupt.edu.cn; cyfeng@bupt.edu.cn).

Matteo Zecchin, Sangwoo Park, and Osvaldo Simeone are with the King's Communications, Learning \& Information Processing (KCLIP) lab, Department of Engineering, King's College London, London WC2R 2LS, U.K. (e-mail: matteo.1.zecchin@kcl.ac.uk; sangwoo.park@kcl.ac.uk; osvaldo.simeone@kcl.ac.uk).
}
}

\maketitle

\begin{abstract}
\hspace{0.0001cm}{\color{blue} In this paper, we consider a wireless federated inference scenario in which devices and a server share a pre-trained machine learning model. The devices communicate statistical information about their local data to the server over a common wireless channel, aiming to enhance the quality of the inference decision at the server.} Recent work has introduced \emph{federated conformal prediction} (CP), which leverages devices-to-server communication to improve the \emph{reliability} of the server's decision. With federated CP, devices communicate to the server information about the loss accrued by the shared pre-trained model on the local data, and the server leverages this information to calibrate a \emph{decision interval}, or \emph{set}, so that it is guaranteed to contain the correct answer with a pre-defined target \emph{reliability} level. Previous work assumed noise-free communication, whereby devices can communicate a single real number to the server. In this paper, we study for the first time federated CP in a wireless setting. We introduce a novel protocol, termed \emph{wireless federated conformal prediction} (WFCP), which builds on \emph{type-based multiple access} (TBMA) and on a novel quantile correction strategy. WFCP is proved to provide formal reliability guarantees in terms of coverage of the predicted set produced by the server. Using numerical results, we demonstrate the significant advantages of WFCP against digital implementations of existing federated CP schemes, especially in regimes with limited communication resources and/or large number of devices.
\end{abstract}


\section{Introduction}
\emph{Federation} is a data processing paradigm whereby distributed devices with local, possibly private, data sets collaborate for the purpose of carrying out a shared information processing task without the direct exchange of the local data sets. The main exemplar of federated data processing is \emph{federated learning}, which addresses the task of training a machine learning model. Federated learning has been widely studied in recent years, with research activities ranging from theoretical analyses \cite{wan2021convergence, xing2021federated} to the design of communication protocols \cite{xing2020decentralized, amiri2020machine} and to testbeds \cite{caldas2018leaf, bonawitz2019towards}. This paper focuses on a different federated data processing task, namely \emph{federated inference}, with the goal of leveraging collaboration across devices to ensure \emph{reliable decision-making}.

\subsection{Federated Reliable Inference}\label{Federated_inference}
\begin{figure}[t]
    \centering
    {\includegraphics[width = 0.45\textwidth]{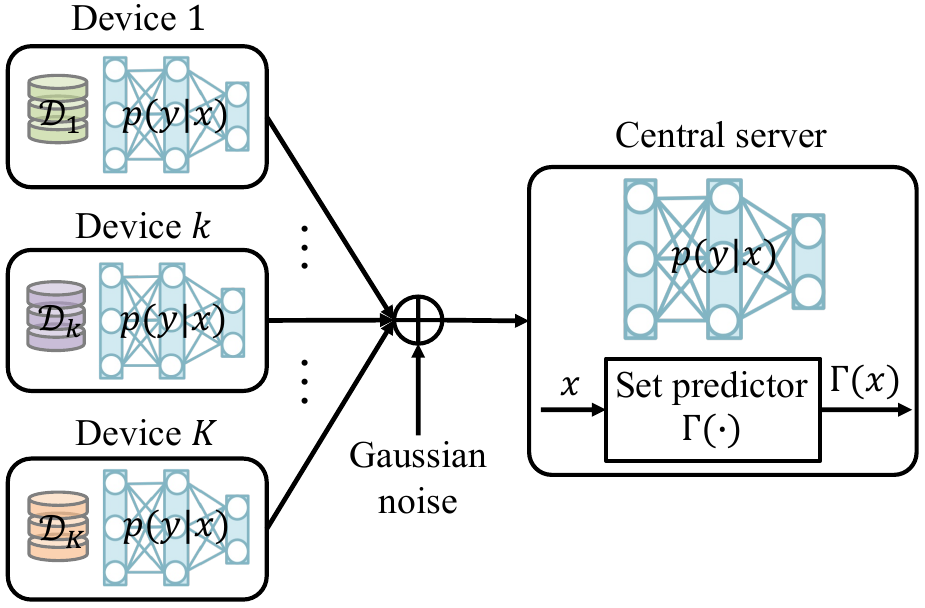}}
    \caption{Illustration of the wireless reliable federated inference problem under study: A \emph{pre-trained} machine learning model $p(y|x)$ is available at devices and a server. The server wishes to make a reliable prediction on a test input $x$, which is not available at the devices. Following the CP framework, the prediction takes the form of a subset $\Gamma(x)$ of the label space $\mathcal{Y}$. The goal is to ensure that the predicted set $\Gamma(x)$ contains the true label with probability no smaller than a target reliability level $1-\alpha$ (see (\ref{coverage})). To this end,  each device $k$ communicates information about the local data set $\mathcal{D}_k$ to the server over a noisy shared channel. This information is then used at the server not to update the model $p(y|x)$ but rather to calibrate the prediction $\Gamma(x)$, ensuring the reliability condition (\ref{coverage}).
    }
    \label{Fig_setting}
\end{figure}

As illustrated in Fig. \ref{Fig_setting},  we study a federated inference setting in which devices and a server share a \emph{pre-trained} machine learning model. The model may have been obtained through a previous phase of federated learning, or it may have been downloaded from a repository of existing models trained in any other arbitrary manner. The server wishes to make an inference on a new input using the model. Devices have access to data, previously not used for training, and can communicate to the server over wireless channels. The devices do not have access to the new input.

{\color{blue}Federated inference has applications in fields as diverse as distributed healthcare platforms \cite{xu2021federated}, Internet-of-Things systems \cite{imteaj2021survey}, and autonomous vehicle networks \cite{zeng2022federated}. As an example, consider a personal healthcare application in which wearable devices collect information about the respective users, while sharing the same pre-trained model. A server in the cloud may wish to draw some conclusions about input data uploaded by another user. Despite not having access to the new input, devices can convey information to the server so as to enhance the quality of inference at the server.}
 
For this setting, recent work has introduced \emph{federated conformal prediction} (CP). {\color{blue}In CP, decisions are in the form of an \textit{interval} or \textit{set}, of possible output values that is guaranteed to contain the correct answer with a pre-defined target \textit{reliability} level. Federated CP leverages devices-to-server communication to support reliable decision-making at the server \cite{FedCP-QQ}. Specifically, devices communicate to the server information about the performance accrued by the shared pre-trained model on the local data.}

Intuitively, this information provides a yardstick with which the server can gauge the plausibility of each value of the output variable for the given input. For instance, if the model obtains a loss no larger than some value $\ell$ on 90\% of the data points at the devices, then the server may safely exclude from the predicted interval/set all output values to which the model assigns a loss larger than $\ell$, as long as it wishes to guarantee a 90\% reliability level. In other words, the server leverages information received from the devices to \emph{calibrate} its decision interval/set.

Previous work \cite{FedCP-QQ} assumed \emph{noise-free} communication, whereby devices can communicate a single real number to the server. Specifically, reference \cite{FedCP-QQ} proposed a \emph{quantile-of-quantile} (QQ) scheme, referred to as \emph{FedCP-QQ}, whereby each device computes and communicates a pre-determined quantile of the local losses. In this paper, we study for the first time federated CP in a wireless setting.

\subsection{Wireless Federated Conformal Prediction}
Even with a perfect transmission of local quantiles, the performance of FedCP-QQ is inherently limited. In fact, for a target reliability level of, say, 90\%, ideally, the server would need to know the 90-percentile of the losses obtained by the pre-trained model across \emph{all} devices. However, the quantile-of-quantiles targeted by FedCP-QQ provides a generally inaccurate estimate of the overall quantile, particularly when the number of devices is large. Furthermore, a direct implementation of  FedCP-QQ \cite{FedCP-QQ} on a wireless channel would require the transmission of quantized local quantiles, requiring a bandwidth that increases proportionally to the number of available devices.

\begin{figure}[t]
    \centering
    {\includegraphics[width = 0.45\textwidth]{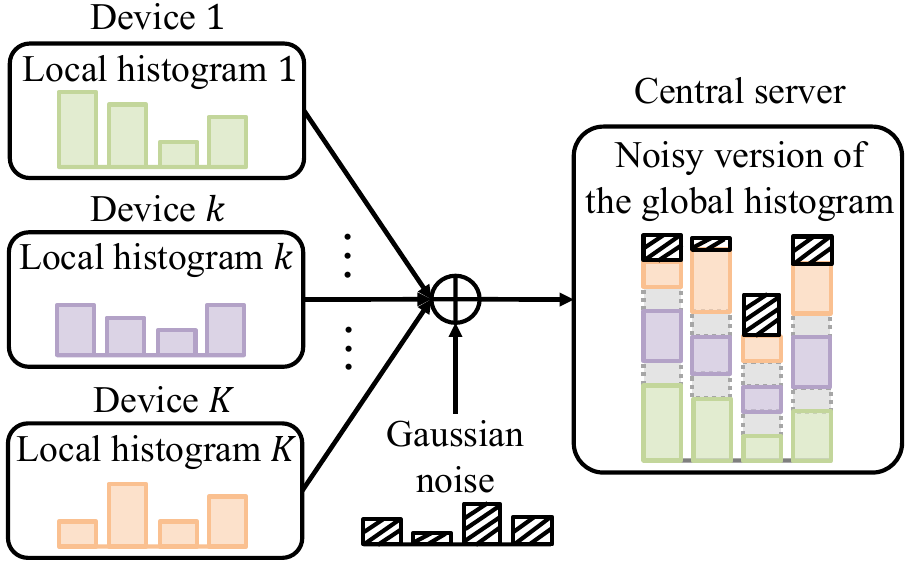}}
    \caption{TBMA enables the estimate of the global histogram of discrete scalar data available across all devices. To this end,  orthogonal codewords are assigned to each histogram bin. All devices transmit simultaneously their individual local histograms over a shared wireless channel by allocating to each codeword a power proportional to the corresponding bin probability. This way, the server can obtain a noisy estimate of the global histogram thanks to the superposition of the signals received for each orthogonal codeword.}
    \label{histogram_illus}
\end{figure}

In this paper, we introduce a novel protocol, termed \emph{wireless federated conformal prediction} (WFCP), which addresses these shortcomings by building on \emph{type-based multiple access} (TBMA) \cite{Tong_TBMA, Liu_TBMA} and on a novel quantile correction scheme. TBMA is a multiple access scheme that aims at recovering aggregated statistics, rather than individual messages. In particular, it can be used to support the estimate of the \emph{histogram} of data available across the devices at the server. To explain it, assume that each device has scalar, quantized, data with a given, generally different,  histogram.  As illustrated in Fig. \ref{histogram_illus}, the goal of the server is to estimate the average histogram across all devices, i.e., the histogram of the data available at all devices, without having to separately estimate the histograms of all devices. 

To accomplish this objective, in TBMA, each histogram bin is assigned an orthogonal codeword. Devices divide their transmission energy across a number of orthogonal codewords in such a way that more energy is allocated to codewords corresponding to histogram bins with a larger number of data points. Allowing for all devices to transmit simultaneously, by collecting the energy received in each bin, the server can estimate the global histogram thanks to the superposition property of wireless communications.

By adopting TBMA as the communication protocol, the proposed WFCP scheme allows the server to estimate the histogram of the losses accrued by the pre-trained model across all the devices. This estimate is then used to estimate the desired global quantile. Importantly, the bandwidth required by TBMA scales only with the resolution of the quantization, i.e., with the number of bins, and not with the number of devices.

The main technical challenge in the design of WFCP is how to ensure reliability -- that is, the condition that the predicted interval/set at the server contains the true output with the desired reliability level. This challenge arises from the fact that the estimate of the global histogram of the losses, and hence of its quantile, is inherently noisy (see Fig. \ref{histogram_illus}). WFCP addresses this problem by proposing a novel quantile correction method that is proved to guarantee reliability.

\subsection{Related Work}
We now provide a brief review of related work by focusing first on federated CP protocols and then on TBMA.

\emph{Federated CP}. Prior to the introduction of the FedCP-QQ scheme \cite{FedCP-QQ} reviewed in Sec. \ref{Federated_inference}, reference \cite{FedCP-Avg} initially applied CP in federated settings, aiming to provide distribution-free, set-valued predictions with reliability guarantees. In \cite{FedCP-Avg}, each device calculates a quantile of its local losses, and the server aggregates these quantiles from all devices to form an average. However, applying CP with the averaged quantile does not guarantee reliability. To address these limitations, FedCP-QQ \cite{FedCP-QQ} was proposed whereby a QQ estimator is used in lieu of an average of quantiles, re-establishing formal reliability guarantees for federated CP.

Federated CP has been further generalized to address settings with statistical heterogeneity across the data available at the devices. In \cite{lu2023federated}, the authors proposed an approach that ensures that the set predictor is well calibrated with respect to a specific mixture of distributions of the devices' local data. To reduce the communication overhead, they proposed to apply distributed quantile estimation methods \cite{luo2016quantiles, dunning2021t} to acquire an approximate quantile. Due to the imperfect estimate of the quantile, reliability guarantees are only proved under additional assumptions on the quality of the estimation error.
In parallel, reference \cite{plassier2023conformal} studied a related setting with label distribution shifts among devices by generalizing the weighted quantile computation scheme proposed in  \cite{tibshirani2019conformal} for centralized CP.
In particular, by noting that a quantile can be obtained as the minimizer of the pinball loss \cite{plassier2023conformal}, the authors applied a gradient-based approach to jointly estimate a quantile from distributed devices.
Accordingly, unlike the setting studied in this paper and in previous work, which assumes one-shot, or embarrassingly parallel, protocols,  the scheme in \cite{plassier2023conformal} is iterative, requiring multiple communication rounds.

All existing federated CP techniques do not consider the influence of noise on the communication channel. This paper aims to address this knowledge gap by investigating the problem of wireless federated CP and by focusing on the impact of channel noise on quantile estimation and, in turn, on model calibration. Unlike \cite{lu2023federated, plassier2023conformal}, as in \cite{FedCP-QQ}, we target formal reliability guarantees without additional assumptions on the quality of the quantile estimates. Furthermore, as in reference \cite{FedCP-QQ}, we focus on statistically homogeneous data across devices.

\emph{TBMA}. The pioneering papers \cite{Tong_TBMA, Liu_TBMA} introduced TBMA, whereby orthogonal codewords are assigned to different measurement values across multiple devices and a variant of the maximum likelihood estimator is devised to accomplish single parameter estimation. In reference \cite{multi_cell_FRAN}, TBMA was applied to estimate multiple correlated parameters in a multi-cell set-up by leveraging in-cell orthogonal TBMA and inter-cell non-orthogonal frequency reuse strategy under centralized and decentralized decoding settings. Furthermore, papers \cite{single_cell_codebook,multi_cell_codebook} developed a non-orthogonal variant of TBMA for multi-valued event detection in random access scenarios. Based on the assumption of sparse user activity, Bayesian approximate message passing estimators were designed for a single-cell \cite{single_cell_codebook} and a multi-cell fog-radio access network \cite{multi_cell_codebook} respectively. Reference \cite{zhu2023information} proposed an end-to-end design of TBMA protocols, whereby the information bottleneck principle is adopted as the criterion to jointly optimize the codebook and neural network-based estimator under unknown source and channel statistics. None of these papers provide any insights into the key problem of using TBMA for reliable quantile estimation.

\subsection{Contributions and Organization}
In this paper, we introduce WFCP, the first wireless protocol for the implementation of federated CP. {\color{blue}At the core of the approach is the integration of TBMA for the estimation of the global histogram of data across all devices and of a novel mechanism to ensure formal reliability guarantees by properly compensating for the errors caused by wireless channels on the TBMA outputs.} The main contributions of this paper are summarized as follows.
\begin{itemize}
  \item We first review conventional centralized CP, which assumes that all data is available at the server \cite{vovk2005algorithmic}. Then, we introduce a digital communication framework in order to apply the state-of-the-art federated CP scheme, FedCP-QQ \cite{FedCP-QQ}, to wireless systems, which will serve as a baseline scheme for WFCP. To this end, we assume that all the devices orthogonally share the available channel uses via time-division multiple access (TDMA), and that the server can detect and discard the received erroneous local quantile to implement the QQ estimator over the correctly received quantiles.
  \item We propose WFCP, a novel protocol based on TBMA that hinges on a carefully selected quantile threshold that accounts for the presence of channel noise.
  \item We provide a rigorous analysis of the reliability performance of WFCP, proving that it can achieve any target reliability level. The analysis also provides guidelines on the choice of important design parameters such as the number of quantization levels.
  \item Simulation results demonstrate the advantage of the proposed WFCP scheme over existing strategies, especially in the presence of limited communication resources and/or large number of devices. 
\end{itemize}

The remainder of this paper is organized as follows. In Sec. \ref{setting}, we describe the setting and define the problem. Sec. \ref{background_CP} presents the general framework of conventional CP, and introduces a quantized version for future reference. Sec. \ref{digital_CP} reviews the FedCP-QQ scheme \cite{FedCP-QQ}, which operates in an ideal noise-free scenario,  and describes a digital wireless implementation of FedCP-QQ. In Sec. \ref{WFCP}, we propose the WFCP scheme, also providing design guidelines and proof of reliability. Sec. \ref{experiment} evaluates the performance of WFCP as compared to benchmarks via experiments, validating the effectiveness of WFCP. Sec. \ref{conclusion} summarizes this paper and points to directions for future work.

\section{Setting and Problem Definition}\label{setting}
\subsection{Setting}
We consider a \emph{wireless federated inference} scenario in which a set of $K$ devices and a central server communicate over a multiple access channel. A \emph{pre-trained} machine learning model is available at both server and devices side. This model may have been previously trained using federated learning \cite{mcmahan2017communication}. As in \cite{FedCP-QQ}, we focus on the problem of \emph{reliable} collaborative, or federated, inference using a fixed model along with communication between devices and server. In this setting, communication is not used to optimize the machine learning model, as in federated learning, but rather to ensure a higher level of reliability for an inference decision produced by the server on a new input available only at the server. As we will detail, thanks to communication, devices can help the server \emph{calibrate} its decision, enhancing the server's estimate of the corresponding uncertainty.

To elaborate, we focus on a classification problem with $C$ classes, which are labelled by elements in set $\mathcal{Y}=\{1,2,\ldots,C\}$. Given any input $x\in\mathcal{X}$, the predictive model produces a conditional probability distribution $p(y|x)$ over the labels $y\in\mathcal{Y}$. Probability $p(y|x)$ is typically interpreted as a measure of the \emph{confidence} that the model has in label $y$ being the correct one. Conventionally, a decision $y^*$ is obtained by selecting the label on which the model has maximum confidence, i.e.,
\begin{equation}\label{point_prediction}
    y^*=\arg\max_{y\in\mathcal{Y}}p(y|x).
\end{equation}

The corresponding confidence level $p(y^*|x)$ produced by the model should ideally provide an indication of the true accuracy of the decision $ y^*$. However, it is well known that machine learning models tend to be overconfident \cite{bai2021don,guo2017calibration,cohen2022calibrating}, and hence systems using the decision $y^*$ produced by the model cannot trust the confidence level $p(y^*|x)$ to provide a reliable measure of the reliability of the decision.

In this paper, we are interested in producing decisions that provide trustworthy measures of uncertainty. To this end, following the CP framework \cite{vovk2005algorithmic}, our goal is to ensure that, based on a generally \emph{unreliable} model $p(y|x)$ and on communication with the devices, the server outputs \emph{set-valued} decisions for any new input $x$ with \emph{formal reliability guarantees}.

To explain, we define a mapping from an input $x$ to a subset of the label space as $\Gamma(x)\subseteq \mathcal{Y}$. Given a new input $x$ available only at the server, as well as the model output distribution $\{p(y|x)\}_{y\in\mathcal{Y}}$, along with information received from the devices, the server aims at producing a set-valued decision $\Gamma(x)$ with reliability guarantees. For a \emph{target reliability level} $1-\alpha$, with $\alpha\in [0,1]$, a set decision is said to be \emph{reliable} if the set contains the true label $y$ with probability at least $1-\alpha$, i.e., if the inequality
\begin{equation}\label{coverage}
    \Pr(y \in \Gamma(x))\geq 1-\alpha
\end{equation}
holds. The probability in (\ref{coverage}) is evaluated with respect to the randomness of data generation and communications, as we discuss in the next subsections.

Before detailing the role of communications, we observe that the reliability requirement (\ref{coverage}) can be trivially met by choosing as a set decision the set of all possible labels, i.e., $\Gamma(x)=\mathcal{Y}$, irrespective of the input $x$. This set predictor, while reliable, would be completely uninformative. It is hence also important to evaluate the performance of the set decision on the basis of the average size of its prediction. This is known as the \emph{inefficiency} of the predictor, which is defined as
\begin{equation}\label{eq:ineff}
    \mathbb{E}\left[\left|\Gamma(x)\right|\right],
\end{equation}
where $|\cdot|$ represents the cardinality of the argument set. The expectation in (\ref{eq:ineff}) is evaluated with respect to the randomness of both data and communications, as in \eqref{coverage}.

\subsection{Data Model}
As mentioned in the previous subsection, we assume that the model $p(y|x)$ is pre-trained using an arbitrary training technique and an arbitrary data set. Accordingly, we do not concern ourselves with the training data and with the training process in this paper. That said, the CP procedure requires a data set -- distinct from training data -- that is used to calibrate model $p(y|x)$ so as to obtain a reliable set-valued prediction in the sense of condition (\ref{coverage}) for some input $x$. In conventional CP, such data set, known as \emph{calibration data set}, is directly available at the decision-maker holding the model and the input $x$. In this paper, as in \cite{FedCP-QQ}, we assume that, instead, calibration data sets are only present at the devices. By communicating information about such data to the server, the devices can facilitate the implementation of CP-based mechanisms to produce reliable estimates $\Gamma(x)$ that satisfy the inequality \eqref{coverage} for a desired reliability level $1-\alpha$. We will explain how CP works in the next section.

We assume that there are a total of $N$ \emph{calibration data points}, denoted as $\mathcal{D}=\left\{(x_i,y_i)\right\}_{i=1}^N$, which are equally split across all $K$ devices. Accordingly, each device $k$ stores $N_d=N/K$ calibration data points, denoted as $\mathcal{D}_k = \left\{(x_{i,k}, y_{i,k})\right\}_{i=1}^{N_d}$. The union of all disjoint sets of data points $\mathcal{D}_k$ across all $K$ devices recovers the overall calibration data set, i.e., $\mathcal{D} =\mathcal{D}_1\cup\cdots\cup\mathcal{D}_K $. Following the standard machine learning model, all calibration data points in data set $\mathcal{D}$ are assumed to be generated i.i.d. from some unknown distribution $p^*(x,y)$. {\color{blue}The generalization to the case of data heterogeneity and different data set sizes is discussed in Sec. \ref{conclusion}.}

Furthermore, we denote a generic \emph{test data point} as $(x, y)$, which is also generated from distribution $p^*(x,y)$, independently from the calibration data. As explained, the input $x$ of the test pair is known only to the server, while the corresponding label $y$ is unknown and must be predicted by the server.

\subsection{Communication Model}\label{subsec:protocols_tdma_tbma}
In prior work on federated inference via CP \cite{FedCP-Avg, FedCP-QQ,lu2023federated,plassier2023conformal}, the communication channels between devices and server were assumed to be ideal. In contrast, in this work, we study the more challenging scenario in which devices are connected to the server via noisy channels. Accordingly, we will refer to this problem as \emph{wireless reliable federated inference}.

Specifically, the $K$ devices communicate with the server over a shared fading multiple access channel using $T$ channel uses, or symbol periods. As we will detail below, we consider two types of protocols, namely orthogonal-access systems in which each device uses distinct subsets of channel uses, and non-orthogonal protocols in which devices are simultaneously active on all channel uses.

We assume an average per-symbol power constraint $P$ for each device $k$. Accordingly, denoting the per-symbol power of the channel noise as $N_0$, we define the signal-to-noise ratio (SNR) as 
\begin{align}\label{eq:def_SNR}
    \text{SNR} = \frac{P}{N_0}.
\end{align}

{\color{blue}
\subsubsection{Orthogonal Multiple Access} Time-division multiple access (TDMA) is a conventional orthogonal multiple access scheme that assigns distinct subsets of the $T$ channel uses to the $K$ devices. In this paper, we focus on equal allocations whereby all devices are assigned $\lfloor T/K\rfloor$ channel uses. Accordingly, in symbol period $t$ assigned to device $k$, the received signal at the central server can be expressed as
\begin{equation}\label{received_signal}
    v_t = h_ku_{k,t} + z_t,
\end{equation}
where $h_k$ is the fading coefficient for device $k$, $u_{k,t}$ is the symbol transmitted by device $k$ at time $t$, and $ z_t\sim \mathcal{N}(0,N_0)$ is the channel noise. We focus on a real-valued channel model, corresponding for instance to the in-phase or quadrature component of a passband channel. The fading channels $h_k$ are assumed to be independent random variables across user index $k$, and they are constant within the $T$ channel uses.
We assume perfect channel state information (CSI) is available to both the devices and the server, and thus, without loss of generality, we constrain the channel coefficients $h_k$ to be non-negative.

\subsubsection{Non-Orthogonal Multiple Access} As we will discuss in Sec. \ref{WFCP}, the proposed scheme relies on TBMA, which is a form of non-orthogonal multiple access protocol. In general, in non-orthogonal protocols, a subset $\mathcal{K}_a\subseteq\left\{1,\ldots,K\right\}$ of the $K$ devices transmit concurrently in each symbol period $t$. Accordingly, the signal received at the server in period $t$ can be written as
\begin{equation}\label{received_signal2}
    v_t = \sum_{k\in\mathcal{K}_a}h_k u_{k,t} + z_t,
\end{equation}
with the same definition given above for orthogonal protocols.}

\section{Background on Conformal Prediction}\label{background_CP}
In this section, we provide a brief primer on CP in order to set the necessary background required by benchmarks and proposed schemes for the problem of wireless reliable federated inference described in the previous section. The presentation also includes discussions about the impact of quantization on the performance of CP, which is not covered in standard references on CP. Unlike the federated setting of interest in this work, conventional CP applies to a centralized scenario with a server holding all the available calibration data, which will be assumed throughout this section.

\subsection{Validation-Based Conformal Prediction}
We focus on a practical variant of CP, known as split, inductive, or \emph{validation-based} CP, that operates on a pre-trained model $p(y|x)$ \cite{vovk2005algorithmic, angelopoulos2021gentle, barber2021predictive}. Given a new input $x$, the goal of CP is to produce a set predictor $\Gamma(x)\in\mathcal{Y}$ with the property of satisfying the reliability condition (\ref{coverage}) for some pre-determined target reliability level $1-\alpha$. To this end, the server is given access not only to the model $p(y|x)$ and to a test input $x$, but also to a \emph{calibration data set } $\mathcal{D}=\left\{(x_i,y_i)\right\}_{i=1}^N$  consisting of $N$ data points. As in the previous section, the $N$ calibration data points and the test data pair $(x,y)$ are assumed to be i.i.d. according to an unknown distribution $p^*(x,y)$. The probability in \eqref{coverage} is evaluated with respect to the joint distribution of calibration and test data.
 
In the centralized setting under study here, the server builds the set predictor $\Gamma(x)$ using the test input $x$, the calibration data, and the model $p(y|x)$. Note that the true label $y$ is not known at the server, since it is the subject of the inference process.

To this end, we introduce the \emph{nonconformity (NC) score function} 
\begin{equation}\label{NC score}
    s(x,y) = 1 - p(y|x).
\end{equation}
The NC score is a measure of the loss of the model $p(y|x)$ on the data point $(x,y)$. In fact, a large value indicates that the model assigns a low probability to example $(x,y)$. Other NC scores are also possible \cite{romano2020classification, wang2022probabilistic, teng2022predictive, chen2023spikecp}, and the methodology developed in this paper applies more broadly to any scores, as long as they are non-negative and upper bounded, i.e.,
\begin{equation}\label{eq:NCbounds}
    0\leq s(x,y)\leq 1.
\end{equation}
Note that the upper bound is set to 1 without loss of generality since one can always re-scale a bounded NC score to fit in the range (\ref{eq:NCbounds}).

CP includes in the predicted set $\Gamma(x)$ all labels $y\in \mathcal{Y}$ with a NC score smaller than a given threshold $s_\alpha$, i.e.,
\begin{equation}\label{org set predictor1}
    \Gamma_{\alpha}(x) = \left\{y\in\mathcal{Y}: {s}(x,y)\leq s_\alpha\right\}.
\end{equation}
As we discuss next, the threshold $s_\alpha$ is determined based on the target reliability level $1-\alpha$ in the reliability constraint (\ref{coverage}). Dependence on parameter $\alpha$ is accordingly added in the subscript of the set predictor $\Gamma_{\alpha}(x)$.

\subsection{Evaluation of the Threshold}
To evaluate the threshold $s_\alpha$, the server computes the NC scores \eqref{NC score} for all the $N$ calibration data points, obtaining the collection $\mathcal{S}\triangleq\left\{s(x_i,y_i)\right\}_{i=1}^N$ of NC scores. Note that multiple data points may have the same NC score, which is accordingly counted multiple times. Then, the server sets the threshold $s_\alpha$ to be approximately equal to the $\lceil(1-\alpha)N\rceil$-th smallest NC score in the set $\mathcal{S}$ (counting possible repetitions). Intuitively, as the reliability level $1-\alpha$ increases, so does the threshold $s_\alpha$, ensuring that the predicted set (\ref{org set predictor1}) includes a larger number of labels.

To formalize the operation of CP, let us introduce a function that, given a set $\mathcal{S}$, produces the $\lceil (1-\alpha)(N+1)\rceil$-th smallest value in the set. Note that the smallest value is evaluated with respect to a set with cardinality $N+1$, and not $N$, as required by CP (see, e.g., \cite{barber2021predictive}). For any given collection of real numbers $\mathcal{S}=\{s_1,...,s_{N}\}$ with possible repetitions, we denote as $s_{(1)}\leq s_{(2)}\ldots\leq s_{(N)}$ the sorted values in ascending order. Ties are broken arbitrarily. Then, the desired function is defined as
\begin{align}\label{quantile_function}
    Q_{1-\alpha}\left(\mathcal{S}\right) \triangleq
    \begin{cases}
        s_{(\lceil (1-\alpha)(N+1)\rceil)} & \text{if $\alpha \geq 1/(N+1)$}, \\
        1 & \text{otherwise},
    \end{cases}
\end{align}
where $\lceil \cdot \rceil$ denotes the ceiling operation. Accordingly, function $Q_{1-\alpha}\left(\mathcal{S}\right)$ returns the $\lceil (1-\alpha)(N+1)\rceil$-th smallest value in the set as long as $\lceil (1-\alpha)(N+1)\rceil\leq N$, or equivalently $\alpha \geq 1/(N+1)$; while returning the maximum value 1 otherwise.

The value $Q_{1-\alpha}\left(\mathcal{S}\right)$ can also be interpreted as the $(1-\alpha)(N+1)/N$-\emph{quantile} of the empirical distribution of the entries of set $\mathcal{S}$. In fact, the $(1-\alpha)(N+1)/N$-quantile of the empirical distribution is, by definition, the smallest number in the set $\mathcal{S}$ that is at least as large as a fraction $(1-\alpha)(N+1)/N$ of the elements in $\mathcal{S}$.

With function $Q_{1-\alpha}(\cdot)$, the CP set predictor (\ref{org set predictor1}) can be succinctly expressed as
\begin{equation}\label{org set predictor}
    \Gamma_{\alpha}(x) = \left\{y\in\mathcal{Y}: s(x,y)\leq s^{\textrm{CP}}_\alpha\triangleq Q_{1-\alpha}\left(\mathcal{S}\right)\right\}.
\end{equation}
As mentioned, it can be proved that the prediction set $\Gamma_{\alpha}(x)$ in \eqref{org set predictor} satisfies the reliability condition \eqref{coverage}, irrespective of the accuracy of the underlying model $p(y|x)$ and of the ground-truth distribution $p^*(x,y)$ of the data \cite{vovk2005algorithmic, lei2014distribution}.

\subsection{Quantized Conformal Prediction}\label{quantized CP}
As discussed in the previous section, in this paper, we are concerned with decentralized settings in which calibration data is not available at the server. In such a setting, communication between devices holding the calibration data and the server is limited by the available transmission resources. As a step in the direction of accounting for limitations arising from finite communication capacity, in this subsection, we discuss a \emph{centralized} CP setting in which NC scores used to evaluate the threshold $s^{\text{CP}}_\alpha$ in (\ref{org set predictor}) are constrained to take a discrete finite set of values.

To this end, we adopt a \emph{uniform scalar quantizer} in which the range $[0,1]$ of possible values for the NC score, by assumption (\ref{eq:NCbounds}), is divided into $M$ equal intervals $[S_0,S_1], (S_1,S_2],\ldots (S_{M-1},S_M]$ with $S_0=0$ and $S_M=1$.
{\color{blue}Given an input NC score $s\in [0,1]$, the quantized output $q(s)$ equals the upper value $S_m$ of the interval $(S_{m-1},S_m]$ containing score $s$. Accordingly, the quantization function is defined as 
\begin{equation}\label{quantizer}
    q \left(s\right)\triangleq
    \begin{cases}
        S_1 & \text{$s\in[S_0,S_1]$},\\
        S_m & \text{$s\in(S_{m-1},S_m]$}~\text{for}~m=2,\ldots,M.
    \end{cases}
\end{equation}
It is also possible to include a non-linear monotonically increasing transformation of the NC scores prior to the application of \eqref{quantizer} as in conventional companding.
}

Suppose now that the server has access to the set of quantized NC scores $\mathcal{S}^{q}\triangleq\left\{q(s(x_i,y_i))\right\}_{i=1}^N$. Following the CP procedure, we define the set predictor as
\begin{equation}\label{set_predictor_p}
    \Gamma^q_{\alpha}(x) = \{y \in \mathcal{Y}: q(s(x, y)) \leq s^{q-\text{CP}}_{\alpha} \triangleq Q_{1-\alpha}(\mathcal{S}^q) \},
\end{equation}
that is, as the set of labels $y\in\mathcal{Y}$ whose quantized NC scores $q(s(x, y))$ are no larger than the $\lceil(1-\alpha)(N+1)\rceil$-th smallest NC score, $Q_{1-\alpha}(\mathcal{S}^q)$, in the calibration set.

Since any function of the input-output pair $(x,y)$ is a valid NC score, so is the quantized value $q(s(x,y))$. Therefore, the quantized predicted set $ \Gamma^q_{\alpha}(x)$ satisfies the reliability condition \eqref{coverage}.  However, one should generally expect that, due to information loss caused by quantization, the size of the predicted set $ \Gamma^q_{\alpha}(x)$ is generally larger than that of the predicted set $ \Gamma_{\alpha}(x)$ obtained from the original NC score function $s(x,y)$.

\subsection{Quantized Conformal Prediction via Empirical Quantiles}\label{quantile by distribution}
In this subsection, we make the observation that the threshold $s^{q-\text{CP}}_{\alpha} = Q_{1-\alpha}(\mathcal{S}^q)$ used in the set predictor \eqref{set_predictor_p} can be expressed in terms of the empirical distribution of the quantized NC scores in set $\mathcal{S}^q$. More precisely, it can be evaluated, approximately, as the $(1-\alpha)$-quantile of the empirical distribution. This fact will be instrumental in the design of the proposed federated inference protocol in Sec. \ref{WFCP}.

To elaborate, let us define as $p_m\in[0,1]$ the fraction of quantized NC scores equal to $S_m$ in the set of quantized NC scores $\mathcal{S}^q$, i.e.,
\begin{align}
    p_{m}=\frac{1}{N}\sum^{N}_{i=1}\mathds{1}\{m_{i}=m\},
\end{align}
where $m_i$ is the index of the quantized $i$-th NC score, i.e., $S_{m_i}=q(s(x_i,y_i))$. We collect all $M$ fractions into the vector
\begin{equation}\label{eq:boldp}
    \boldsymbol{p}=[p_1,\ldots,p_M]^{\mathrm{T}},
\end{equation}
which satisfies the equality  $\sum_{m=1}^M p_m=1$. As we have discussed, CP relies on the evaluation of the $\lceil(1-\alpha)(N+1)\rceil$-th smallest element in the set $\mathcal{S}^q$, i.e., $ Q_{1-\alpha}(\mathcal{S}^q)$. To evaluate this quantity, we modify the empirical distribution of the quantized NC scores by adding a fictitious $(N+1)$-th NC score equal to the maximum value. This yields the empirical distribution vector 
\begin{equation}\label{p+}
    \boldsymbol{p}^{+}=\frac{N}{N+1}\boldsymbol{p} + \left[0,\ldots,\frac{1}{N+1}\right]^{\mathrm{T}}.
\end{equation}

With this definition, the $\lceil(1-\alpha)(N+1)\rceil$-th smallest element in set $\mathcal{S}^q$ can be obtained as the quantization level $S_{m_{\alpha}(\boldsymbol{p}^+)}$, where the index $m_{\alpha}(\boldsymbol{p}^+)$ is obtained by evaluating the $(1-\alpha)$-\emph{quantile} of the empirical distribution $\boldsymbol{p}^{+}$, i.e.,
\begin{equation}\label{index}\small
    \begin{aligned}
        m_{\alpha} & (\boldsymbol{p}^{+}) \\
        =& \min \Big\{m \in \{1,\ldots,M\}: p^{+}_1 + \cdots + p^{+}_m \geq 1-\alpha \Big\}.
    \end{aligned}
\end{equation}

\section{Digital Wireless Federated Conformal Prediction}\label{digital_CP}
While the conventional CP scheme reviewed in the previous section assumes that predictive model and calibration data are both present at the server, in the wireless reliable federated inference setting as explained in Sec. \ref{setting}, calibration data are only available at the devices. In this section, we first review the FedCP-QQ scheme proposed in \cite{FedCP-QQ}, which addresses this problem by assuming noiseless links from devices to server that can support the noiseless transmission of a single real number from each device. Then, as a benchmark, we describe a direct digital wireless implementation of FedCP-QQ that accounts for the presence of noisy channels between devices and server.

\subsection{Federated Conformal Prediction with Noiseless Communications}\label{FedCPQQ}
The FedCP-QQ scheme introduced in \cite{FedCP-QQ} is based on the \emph{quantile-of-quantiles} (QQ) operation. Accordingly, as we detail next, it sets two probabilities $\alpha_d$ and $\alpha_s$ to identify target quantiles to be computed at devices and server, respectively.

Each device $k$ has access to the \emph{local NC scores} $\mathcal{S}_k= \left\{s(x_{i,k}, y_{i,k})\right\}_{i=1}^{N_d}$. Based on this collection of NC scores, it computes the $(1-\alpha_d)(N_d+1)/N_d$-quantile $Q_{1-\alpha_d}(\mathcal{S}_k)$. This real positive number is then communicated noiselessly to the server.

The server collects all the quantiles $\mathcal{Q}^{1:K}_{1-\alpha_d}=\{Q_{1-\alpha_d}\left(\mathcal{S}_1\right), \ldots, Q_{1-\alpha_d}\left(\mathcal{S}_K\right)\}$ from the $K$ devices, and it evaluates the $(1-\alpha_d)(K+1)/K$-quantile of the $K$ quantiles, i.e.,
\begin{equation}\label{QQ estimator}
    s^{\textrm{QQ}}_\alpha \triangleq Q_{1-\alpha_s}(\mathcal{Q}^{1:K}_{1-\alpha_d}).
\end{equation}

The set predictor of the FedCP-QQ scheme is constructed using the obtained threshold as
\begin{equation}\label{DQQ_set_pre}
    \Gamma_{\alpha_d,\alpha_s}^{\text{QQ}}(x) = \left\{y\in\mathcal{Y}: s(x,y)\leq s^{\textrm{QQ}}_\alpha\right\}.
\end{equation} 
The pair of miscoverage levels $(\alpha_d, \alpha_s)$ must be selected in order to satisfy the coverage condition \eqref{coverage}. To this end, reference \cite{FedCP-QQ} proved the following result.

\begin{theorem}[\textbf{Theorem 3.2 \cite{FedCP-QQ}}]
    \label{theorem_FedCPQQ}
    For any $(\alpha_d,\alpha_s)\in[1/(N_d+1),1)\times[1/(K+1),1)$, the coverage of the set predictor $\Gamma_{\alpha_d,\alpha_s}(x)$ is lower bounded as
    \begin{equation}\label{M_nk}\small
        \begin{aligned}
            &\Pr \left( y \in \Gamma_{\alpha_d,\alpha_s}^{\text{\rm{QQ}}}(x)\right) \\
            &\geq 1 - \frac{1}{N + 1}\sum_{j=k}^K\binom{K}{j}\sum_{I_{1, j}=n}^{N_d}\sum_{I_{1, j}^c=0}^{n-1}\frac{\binom{N_d}{i_1}\cdots{\binom{N_d}{i_m}}}{\binom{N}{i_1+\cdots+i_K}}\triangleq M_{\alpha_d,\alpha_s}
        \end{aligned}
    \end{equation}
    with $n = \lceil (N_d+1)(1-\alpha_d)\rceil$; $k = \lceil (K+1)(1-\alpha_s)\rceil$;  $I_{1, j}=\left\{i_1, \ldots, i_j\right\}$; $I_{1, j}^c=\left\{i_{j+1}, \ldots, i_K\right\}$; and the operation $\sum_{I_{1,j}=n}^{N}$ stands for the cascade of summations that takes into account for all elements in $I_{1,j}$ starting from $n$ up to $N$, i.e., $\sum_{I_{1,j}=n}^{N}= \sum_{i_1=n}^N \sum_{i_2=n}^N \cdots \sum_{i_j=n}^N$.
\end{theorem}

With this result, one can find a pair of miscoverage levels $(\alpha_d,\alpha_s)$ that minimizes the lower bound $M_{\alpha_d, \alpha_s}$ while satisfying the target coverage rate $1-\alpha$. The optimization objective can be formulated as
\begin{equation}\label{optimal_n_k}
    \left(\alpha_d^*, \alpha_s^*\right)\in \arg \min_{\alpha_d, \alpha_s}\left\{M_{\alpha_d,\alpha_s}: M_{\alpha_d,\alpha_s} \geq 1-\alpha\right\}.
\end{equation}
If the solution of \eqref{optimal_n_k} is not unique, it is suggested to find the pairs with the largest value $\alpha_d^*$ and then choose among those the pair with the largest value $\alpha_s^*$. Efficient ways to address this problem are discussed in \cite{FedCP-QQ}, which also covers the more general case in which devices have different data set sizes.

\subsection{Digital Transmission Benchmark}\label{Digital FedCP-QQ}
In this subsection, we propose a digital implementation of the FedCP-QQ scheme, which we refer to as Digital FedCP-QQ or DQQ for short.
A direct implementation of the FedCP-QQ scheme requires every device $k$ to quantize its local quantile $Q_{1-\alpha_d}(\mathcal{S}_k)$ in \eqref{QQ estimator} before transmission in order to meet the capacity constraints on the shared noisy channel to the receiver. To this end, the device $k$  applies the function $q(\cdot)$ defined in \eqref{quantizer} to quantize the local quantile into one of $M$ levels. Then, each device uses conventional digital communications to convey the quantized quantile to the server.

Specifically, to transmit the quantized local quantiles from $K$ devices on the shared channel, we adopt a TDMA protocol whereby, as discussed in Sec. \ref{subsec:protocols_tdma_tbma}, the $K$ devices are assigned $\lfloor T/K\rfloor$ channel uses each.
{\color{blue} Accordingly, the probability of error for each device $k$ with fading coefficient $h_k$ can be closely approximated by the outage probability as \cite{yang2013block, yang2013quasi}
\begin{equation}\label{P_error}
    \epsilon=\Pr\left(\frac{1}{2}\log\left(1+\text{SNR}h_k^2\right)\leq \frac{\log M}{\lfloor T/K\rfloor}\right),
\end{equation}
where the probability is computed with respect to the distribution of the fading channel power $h_k^2$.}

Accordingly, with probability $\epsilon$, the transmission is unsuccessful. Assuming that the server can detect errors, the QQ estimator \eqref{QQ estimator} can be applied on the subset of quantiles that are received correctly. Note that the bound in \eqref{M_nk} should now be evaluated by including only the correctly received quantiles from the devices. {\color{blue} Furthermore, the probability $\alpha_s$ is chosen based on the number of devices whose messages are received without error.}

While the resulting set predictor satisfies the reliability condition (\ref{coverage}) by Theorem 1, the impact of lost quantiles due to channel errors is that of reducing the number of active devices, and hence the amount of calibration data effectively accessible by the server. This, in turn, generally increases the average predicted set size (\ref{eq:ineff}).

\section{Wireless Federated Conformal Prediction} \label{WFCP}
In this section, we introduce the proposed Wireless Federated Conformal Prediction (WFCP) scheme. WFCP implements a novel combination of TBMA and over-the-air computing to communicate the empirical distribution of the quantized NC scores from the {\color{blue}active} devices to the server. Via over-the-air computing, thanks to the superposition property of the multiple access channel \eqref{received_signal2}, the server obtains a noisy and unbiased estimate of the empirical distribution of the NC scores across the {\color{blue}active} devices. Based on this estimate, the server computes an estimate of a global empirical quantile, which is judiciously selected to ensure the coverage property (\ref{coverage}).

Unlike the existing FedCP-QQ scheme reviewed in the previous section, WFCP does not require devices to compute their local quantiles. This local computation, implemented by FedCP-QQ to reduce bandwidth requirements, generally results in a performance loss, since the QQ estimator \eqref{QQ estimator} used by FedCP-QQ cannot recover the global quantile required to implement CP on the overall calibration data set stored across the {\color{blue}active} devices as reviewed in Sec. \ref{background_CP}. This result could only be achieved by communicating separately all the quantized NC scores from the {\color{blue}active} devices to the server. However, for a given transmission reliability level (\ref{P_error}), this transmission would require a number $T$ of channel uses that scale linearly with the number of calibration data points across the {\color{blue}active} devices.

To mitigate this loss, WFCP enables a direct estimate of the global quantile at the server without imposing bandwidth requirements that scale linearly with the number of {\color{blue}active} devices. Rather, the communication requirements of WFCP are only dictated by the precision with which the NC scores are represented for transmission to the server.

In the following, we first detail the transmission protocol based on TBMA adopted by WFCP. Then, we describe the set predictor implemented by WFCP on the basis of the received baseband signal. As it will be detailed, the main challenge in ensuring the reliability condition \eqref{coverage} is the determination of a suitable correction for the quantile estimated based on the noisy received signals. Finally, we provide some discussion on the trade-offs involved in the design choices, and we prove that condition \eqref{coverage} is guaranteed by WFCP.

\subsection{TBMA-Based Communication Protocol} \label{scheme}
{\color{blue}Let us fix a subset $\mathcal{K}_a\subseteq \left\{1,\ldots,K\right\}$ comprising $K_a=|\mathcal{K}_a|$ devices that are active in the given $T$ symbols. As we will detail later in this subsection, this subset depends on the channel gains $\left\{h_k\right\}_{k=1}^K$.}
In WFCP, unlike FedCP-QQ, the {\color{blue}active} devices do not first compute the empirical quantiles of their respective local NC scores. Rather, each {\color{blue}active} device quantizes separately each of the $N_d$ local NC scores using the uniform quantizer \eqref{quantizer} with $M$ levels. We denote as $m_{i,k}\in\{1,\ldots,M\}$ the index of the quantization value produced for the $i$-th NC score at device {\color{blue}$k\in\mathcal{K}_a$}, $s(x_{i,k},y_{i,k})$, i.e.,
\begin{equation}
    S_{m_{i,k}}=q(s(x_{i,k},y_{i,k})).
\end{equation}
Furthermore, in a manner that mirrors the presentation in Sec. \ref{quantile by distribution}, we introduce an $M$-dimensional probability vector that collects the quantized NC scores at device {\color{blue}$k\in\mathcal{K}_a$} as
\begin{equation}\label{eq:pk}
    \boldsymbol{p}_k=[p_{1,k},\dots,p_{M,k}]^{\mathrm{T}},
\end{equation}
where the probability
\begin{align}
    p_{m,k}=\frac{1}{N_d}\sum^{N_d}_{i=1}\mathds{1}\{m_{i,k}=m\}
\end{align}
corresponds to the fraction of NC scores associated to quantization level $m$ at device {\color{blue}$k\in\mathcal{K}_a$}, such that the equality $\sum_{m=1}^M p_{m,k}=1$ holds.

{\color{blue}As an intermediate goal, WFCP obtains an unbiased estimate of the \emph{subsampled global empirical distribution}
\begin{equation} \label{sub_glob_emp}
    \tilde{\boldsymbol{p}}=\frac{1}{K_a}\sum_{k\in\mathcal{K}_a}\boldsymbol{p}_k
\end{equation}
of the quantized NC scores, which is the histogram of the quantized NC scores from the active devices in set $\mathcal{{K}}^a$. The subsampled histogram $\tilde{\boldsymbol{p}}$ equals the \textit{global empirical distribution} $\boldsymbol{p}$ in \eqref{eq:boldp} when all the devices are active, i.e., $K_a=K$.} To this end, we first note that we have the equality
\begin{align}\label{G_distr}
    p_{m}=\frac{1}{K_a}\sum_{k\in\mathcal{K}_a}p_{m,k},
\end{align}
and hence the fraction $ p_{m}$ of NC scores in the $m$-th quantization bin is equal to the corresponding fractions $p_{m,k}$ across the {\color{blue}active devices $k\in\mathcal{K}_a$}. Once such an estimate is available, WFCP can apply the procedure discussed in Sec. \ref{quantile by distribution} in order to mimic the operation of centralized quantized CP. As we will see, this requires a judicious adjustment of the threshold used in evaluating the predicted set \eqref{set_predictor_p}.

To obtain an estimate of distribution {\color{blue}$\tilde{\boldsymbol{p}}$}, WFCP leverages TBMA and over-the-air computing. With TBMA, the devices share a codebook $\boldsymbol{C} = \left[\boldsymbol{c}_1, \ldots, \boldsymbol{c}_M \right] \in \mathbb{R}^{M \times M}$ of $M$ orthogonal codewords, where each codeword $\boldsymbol{c}_m\in\mathbb{R}^{M\times1}$ consists of $M$ real symbols. The number $M$ of channel uses should not be larger than the available number $T$ of symbols. We will discuss the choice of the number of quantization points in Sec. \ref{opt_M}. Assuming that each codeword is normalized to satisfy the energy constraint $\left\|\boldsymbol{c}_m \right\| ^2 = 1$, by the orthogonality of the codewords, we have the equality $\boldsymbol{C}^{\mathrm{T}}\boldsymbol{C}=\mathbf{I}_{M\times M}$. Each codeword $\boldsymbol{c}_m$ is assigned to the $m$-th quantization level $S_m$ for $m\in\{1,\ldots,M\}$.

Accordingly, each {\color{blue}active device $k\in\mathcal{K}_a$} transmits a superposition of the codewords $\boldsymbol{c}_{m_{i,k}}$ that correspond to the $N_d$ quantized NC score  $S_{m_{i,k}}$ for $i=1,...,N_d$. We use the simplified notation $\boldsymbol{c}_{i,k}=\boldsymbol{c}_{m_{i,k}}$. To express the transmitted signal mathematically, let us denote as $\boldsymbol{e}_{i,k}\in\mathbb{B}^{M\times 1}$ the one-hot vector with all zero entries except for a single $1$ at the $m_{i,k}$-th position. The scaled superposition of the codewords transmitted by the device {\color{blue}$k\in\mathcal{K}_a$} can then be written as
{\color{blue}\begin{align}\label{signal}
    \boldsymbol{u}_k = \gamma_k\sum^{N_d}_{i=1}\boldsymbol{c}_{i,k} = \gamma_k\boldsymbol{C} \sum_{i=1}^{N_d}\boldsymbol{e}_{i,k} = \gamma_kN_d\boldsymbol{C}\boldsymbol{p}_k,
\end{align}}where $\gamma_k>0$ is a power control gain at device $k$, and we recall that $\boldsymbol{p}_k$ in (\ref{eq:pk}) is the empirical probability vector of the quantized NC scores at device $k$.

{\color{blue}All the active devices transmit simultaneously on the shared fading channel \eqref{received_signal2}. Accordingly, by \eqref{received_signal2}, thanks to the superposition property of the multiple access channel, the received signal at the server is given by
\begin{equation}\label{received}
    \boldsymbol{v}= \sum_{k\in\mathcal{K}_a} h_k\boldsymbol{u}_k+ \boldsymbol{z} = \boldsymbol{C}N_d\sum_{k\in\mathcal{K}_a}h_k\gamma_k\boldsymbol{p}_k + \boldsymbol{z},
\end{equation}
where $\boldsymbol{z}\in\mathbb{R}^{M\times 1} \sim \mathcal{N}(\boldsymbol{0}, N_0 \mathbf{I})$ is an i.i.d. Gaussian noise vector with zero mean and variance $N_0$, and $\boldsymbol{v}\in\mathbb{R}^{T\times 1}$ collects all received signals within the $T$ channel uses.}

\subsection{\color{blue}Estimate of the Subsampled Global Empirical Distribution}\label{P_usage}
The server wishes to extract an estimate of the {\color{blue}subsampled global empirical distribution $\tilde{\boldsymbol{p}}$} of the quantized NC scores in \eqref{sub_glob_emp} from the received signal \eqref{received}. To this end, matched filtering is applied by left-multiplying the received signal by matrix $\boldsymbol{C}^{\mathrm{T}}$, yielding
{\color{blue}
\begin{equation}\label{matched_filtering}
    \boldsymbol{w} = \boldsymbol{C}^{\mathrm{T}}\boldsymbol{v} = N_d\sum_{k\in\mathcal{K}_a}h_k\gamma_k\boldsymbol{p}_k + \boldsymbol{C}^{\mathrm{T}}\boldsymbol{z}.
\end{equation}
From \eqref{matched_filtering}, the server obtains a weighted sum of the local empirical distribution vectors from the active devices in $\mathcal{K}_a$.

In order to facilitate the recovery of a noisy version of the empirical distribution vector $\tilde{\boldsymbol{p}}$ from \eqref{matched_filtering}, we implement a standard truncated power inversion scheme \cite{elsawy2014stochastic}. Accordingly, we fix a minimum channel power $h_{\text{min}}^2$ such that all users with channel gain power smaller than $h_{\text{min}}^2$ do not transmit. In this way, the active devices in $\mathcal{K}_a$ are selected as all devices $k$ with channel power $h^2_k\geq h^2_{\text{min}}$. For each device $k\in\mathcal{K}_a$, we set the power control coefficients as positive $\gamma_k=\gamma/h_k$ for some scaling factor $\gamma$. Overall, we have
\begin{equation}\label{p_control_para}
    \gamma_k =
    \begin{cases}
        0 & \text{if $h_k^2<h_{\text{min}}^2$},\\
        \gamma/h_k & \text{if $h_k^2\geq h_{\text{min}}^2$}.
    \end{cases}
\end{equation}

The scaling parameters $\gamma$ must satisfy the average per-symbol transmit power constraint $P$. This constraint results in the inequalities
\begin{align}\label{P_constraint}
    \left\| \boldsymbol{u}_k\right\| ^2 = \frac{\gamma^2}{h_k^2}N_d^2\left\| {\boldsymbol{p}}_k\right\| ^2 \leq MP,\quad \text{for}~ k\in\mathcal{K}_a,
\end{align}
We can rewrite \eqref{P_constraint} as
\begin{align}\label{G_ineq}
    \gamma\leq \frac{\sqrt{MP} h_k}{N_d\left\|\boldsymbol{p}_k\right\|}=\frac{\sqrt{MP} h_k}{N_d2^{- H_2({\boldsymbol{p}}_k)}},\quad \text{for}~ k\in\mathcal{K}_a
\end{align}
where $H_2(\boldsymbol{p}_k)$ is the $2$-Rényi entropy \cite{renyi1961measures}
\begin{align}
    H_2(\boldsymbol{p}_k)=-\log_2\left(\sum_{m=1}^{M}p_{m,k}^2\right)
\end{align}
of the empirical probability vector $\boldsymbol{p}_k$. Inequality \eqref{G_ineq} reflects the fact that a more concentrated empirical distribution, with a smaller $2$-Rényi entropy, yields a stricter restriction on the choice of the transmit power. Given that, in general, no prior information is available on the distribution of the NC scores, we set the power scaling factor $\gamma$ by considering the worst-case situation, yielding the choice
\begin{align}\label{gamma}
    \gamma = \frac{\sqrt{MP}h_{\text{min}}}{N_d}.
\end{align}

With \eqref{p_control_para} and \eqref{gamma} in \eqref{matched_filtering}, the matched filtered received signal is given by
\begin{equation}\label{matched_filtering_2}
    \boldsymbol{w} = \sqrt{MP}h_{\text{min}}\sum_{k\in\mathcal{K}_a}\boldsymbol{p}_k + \boldsymbol{C}^{\mathrm{T}}\boldsymbol{z} = \sqrt{MP}h_{\text{min}}K_a\tilde{\boldsymbol{p}} + \boldsymbol{C}^{\mathrm{T}}\boldsymbol{z},
\end{equation}
which is indeed a scaled and noisy version of the empirical probability distribution of the $N_a=K_aN_d$ NC scores from the active devices in $\mathcal{K}_a$.

\subsection{Set Predictor}\label{analysis}
Following the steps presented in the previous subsection, the server recovers the scaled and noisy version $\boldsymbol{w}$ of the empirical distribution $\tilde{\boldsymbol{p}}$ of the $N_a$ calibration NC scores from the active devices in $\mathcal{K}_a$. In the ideal case of no fading and noiseless channels, i.e., with $h_k=1$ for $k=1,\ldots,K$ and $N_0=0$, without power inversion, the server would have access to the empirical distribution $\boldsymbol{p}$ of all the NC scores, and the quantized CP procedure in Sec. \ref{quantile by distribution} could be directly applied to obtain a reliable set predictor.

To address the availability of an estimate of vector $\tilde{\boldsymbol{p}}$, the proposed WFCP preprocesses the scaled and noisy version of the subsampled global empirical distribution \eqref{matched_filtering_2}, and then it computes the $(1-\alpha_c)(N_a+1)/N_a$-quantile of the distribution at a \emph{corrected unreliability level} $\alpha_c<\alpha$, accounting for the presence of channel noise. We will demonstrate that, with a specific choice of the corrected unreliability level $\alpha_c$, the proposed approach preserves the coverage property \eqref{coverage}, with probability now taken also over the channel noise.

First, in order to facilitate the estimate of the subsampled global empirical distribution of the quantized NC scores in \eqref{sub_glob_emp} from the estimate \eqref{matched_filtering_2}, the server carries out two steps: (\emph{i}) it rescales the vector \eqref{matched_filtering_2} by $N_a/(\sqrt{MP}h_{\text{min}}K_a(N_a+1))$; and (\emph{ii}) it adds $1/(N_a+1)$ to the last entry of the received signal vector. Step (\emph{ii}) amounts to the same operation carried out in \eqref{p+} within the centralized quantized CP scheme reviewed in Sec. \ref{quantile by distribution}.

This preprocessing yields the $M\times 1$ vector
\begin{equation}\label{r}
    \boldsymbol{r}= \frac{N_a}{\sqrt{MP}h_{\text{min}}K_a(N_a+1)} \boldsymbol{w} + \left[0,\ldots,\frac{1}{N_a+1}\right]^{\mathrm{T}}=\tilde{\boldsymbol{p}}^++{\tilde{\boldsymbol{z}}},
\end{equation}
where $\tilde{\boldsymbol{p}}^+$ is the empirical distribution vector of the aggregated NC scores from the $K_a$ devices as well as an additional NC score at the maximum quantization level $S_M$, which is equivalent to $\boldsymbol{p}^+$ in \eqref{p+} when all the devices are active, i.e., $K_a=K$, in the case of favorable channel condition; and ${\tilde{\boldsymbol{z}}}\triangleq N_a/(\sqrt{MP}h_{\text{min}}K_a(N_a+1)) \boldsymbol{C}^{\mathrm{T}}\boldsymbol{z}\sim \mathcal{N}(0,\sigma^2 \mathbf{I})$ is the effective noise vector with power
\begin{align}\label{P_eff_noi}
    \sigma^2 &= \frac{N_a^2}{MPh_{\text{min}}^2 K_a^2(N_a+1)^2} N_0 \nonumber\\
    &= \frac{N_d^2}{Mh_{\text{min}}^2\text{SNR}(N_a+1)^2} \approx \frac{1}{Mh_{\text{min}}^2\text{SNR}K_a^2}.
\end{align}

Note that for a fixed number $N_a$ of NC scores, the effective noise power is inversely proportional to the square of the number $K_a$ of active devices. This can be interpreted as a form of coherent gain due to the use of TBMA \cite{Tong_TBMA}. Furthermore, the threshold $h^2_\text{min}$ affects the effective noise power \eqref{P_eff_noi} both directly and through $K_a$ via the power control strategy \eqref{p_control_para}.}


Then, WFCP computes the index $m_{\alpha_c}(\boldsymbol{r})$ corresponding to the $(1-\alpha_c)$-``quantile'' of the noisy distribution $\boldsymbol{r}$. Note that, since the vector $\boldsymbol{r}$ is not normalized, and its entries may be even negative, the index $m_{\alpha_c}(\boldsymbol{r})$ does not correspond to a true quantile in general. To proceed, we define the set
\begin{align}\label{noisy_quantile}
    \mathcal{M}_{\alpha_c}(\boldsymbol{r})&= \Big\{m \in \{1,\ldots,M\}: r_1 + \cdots + r_m \geq 1-\alpha_c \Big\}\nonumber\\
    &= \Big\{ m \in \{1,\ldots,M\}: p_1^+ + \cdots + p_m^+\nonumber\\
    &\quad\quad\quad\quad\geq 1-\alpha_c-(\tilde{z}_1+\cdots+\tilde{z}_m) \Big\}.
\end{align}
Noting that there may be no value $m \in \{1,\ldots,M\}$ that satisfies the inequality in \eqref{noisy_quantile},  we define the index $m_{\alpha_c}(\boldsymbol{r})$ as 
\begin{equation}\label{index_quantile}
    m_{\alpha_c} (\boldsymbol{r})=
    \begin{cases}
	M & \text{if $\mathcal{M}_{\alpha_c}(\boldsymbol{r})=\emptyset$} ,\\
    \min\mathcal{M}_{\alpha_c} (\boldsymbol{r}) & \text{otherwise}.
    \end{cases}
\end{equation}
{\color{blue}In the presence of ideal communications, i.e., with $h_k=1$ for all devices $k$ and with $\sigma^2=0$, the index $ m_{\alpha_c}(\boldsymbol{r})$ corresponds to the index in \eqref{index} computed by the centralized quantized CP with $\alpha_c=\alpha$.}

The index \eqref{index_quantile} is used to define the WFCP set predictor
\begin{align}\label{set_predictor}
    \Gamma_{\alpha_c}^{\text{WFCP}}(x|\boldsymbol{r})= \{y \in \mathcal{Y}: q(s(x, y)) \leq s^{\text{WFCP}}_{\alpha} \triangleq S_{m_{\alpha_c}(\boldsymbol{r})} \}.
\end{align}
{\color{blue}Setting $\alpha_c=\alpha$, the WFCP predicted set \eqref{set_predictor} coincides with the quantized CP predictor \eqref{set_predictor_p} in the presence of ideal communications.}

\subsection{Reliability Analysis} \label{sec:coverage_analysis}
WFCP satisfies the following reliability guarantee.
\begin{theorem}\label{theorem_corr_quan}
Select the corrected unreliability level as
\begin{equation}\label{alphac}
    \alpha_c = \alpha - \frac{\sigma^2 M}{4 \alpha}.
\end{equation}
Then, {\color{blue}for each channel realization,} the WFCP set predictor \eqref{set_predictor} satisfies the reliability guarantee
\begin{equation}\label{cov_bound}
    \Pr\left(y \in \Gamma_{\alpha_c}^{\text{\rm{WFCP}}}(x|\boldsymbol{r})\right)\geq 1-\alpha,
\end{equation}
where the average is taken with respect to the joint distribution of calibration and test data, as well as over the channel noise.
\end{theorem}
\textit{Proof:} The proof is detailed in Appendix \ref{theorem1}.

Theorem \ref{theorem_corr_quan} determines a correction for the threshold level $\alpha_c<\alpha$ in the presence of non-zero Gaussian-noise with effective noise power $\sigma^2$, in order to ensure the satisfaction of the reliability constraint (\ref{coverage}). The correction term $\sigma^2 M/4\alpha$ increases with the effective channel noise power $\sigma^2$ and with the number $M$ of quantization levels.

{\color{blue}Plugging in the definition \eqref{P_eff_noi} of the effective noise power, the correction term can be expressed as
\begin{align}\label{corr_term_1}
    \frac{\sigma^2 M}{4\alpha} = \frac{N_d^2}{4\alpha h_{\text{\rm{min}}}^2 \text{\rm{SNR}}(N_a+1)^2} 
    \approx \frac{1}{4\alpha h_{\text{min}}^2\text{SNR}K_a^2},
\end{align}}which is inversely proportional to the square of the number {\color{blue}$K_a$ of active devices} and to the number $T$ of channel uses available for transmission. The dependence on $K$ is particularly noteworthy: While conventional protocols like DQQ require communication resources in terms of channel uses $T$ and/or SNR, which become increasingly more stringent as $K$ increases, WFCP can benefit from the presence of multiple devices. In particular, as $K$ grows, the correction term in \eqref{corr_term_1} decreases as $1/K^2$, allowing WFCP to set a target reliability level $1-\alpha_c$ that becomes increasingly close to the true target $1-\alpha$.

\subsection{Optimization of the Number of Quantization Levels}\label{opt_M}
The number of quantization levels, $M$, causes an increase in the number of channel uses necessary for the server to recover vector $\boldsymbol{r}$ in \eqref{r}. On the flip side, in the case of ideal communications, a larger value of $M$ generally yields a more informative set predictor thanks to the higher resolution of the NC scores.

{\color{blue}
As explained in Sec. \ref{subsec:protocols_tdma_tbma}, the system has access to $T$ channel uses for transmission from devices to server. A possible choice of the number of quantization levels would be to set $M=T$ which results in high resolution of the NC scores and low effective noise power as indicated in \eqref{P_eff_noi}. As we argue next, however, this may not necessarily be the optimal design since $M$ also influences the correction term in \eqref{alphac}.

With $M<T$, a repetition coding strategy stipulates that the devices repeat their transmission $R\triangleq \lfloor T/M\rfloor$ times where $R$ is the \textit{repetition rate}. The remaining $T-\lfloor T/M\rfloor M$ samples, if any, are not used. With this approach, the effective SNR, upon averaging the matched filter outputs \eqref{matched_filtering_2}, equals
\begin{equation}\label{eff_SNR}
    \text{SNR}_{\text{rep}}=R \cdot\text{SNR} = \lfloor T/M \rfloor\text{SNR}.
\end{equation}

By replacing the SNR in \eqref{P_eff_noi} with \eqref{eff_SNR} obtained through repetition coding, the resulting effective noise power is
\begin{align}\label{sigma_rep}
    \sigma^2_{\text{rep}} &= \frac{N_d^2}{Mh_{\text{min}}^2\text{SNR}_{\text{rep}}(N_a+1)^2} \approx \frac{1}{Th_{\text{min}}^2\text{SNR}K_a^2},
\end{align}
which is approximately independent of the number $M$ of quantization levels. Furthermore, the corresponding correction term in \eqref{corr_term_1} can be accordingly approximated as
\begin{align}\label{corr_term}
    \frac{\sigma_{\text{rep}}^2 M}{4\alpha}
    \approx \frac{M}{4\alpha Th_{\text{min}}^2\text{SNR}K_a^2},
\end{align}
which increases with $M$.

Overall, the choice of the number $M$ of quantization levels entails a tension between improving the resolution of the CP set predictor, which would require increasing the value of $M$, and reducing the correction term in \eqref{corr_term} for a less conservative correction, which calls for a decrease in the value of $M$.}

\begin{figure*}[t]
    \centering
    {
	\includegraphics[width = 0.4\textwidth]{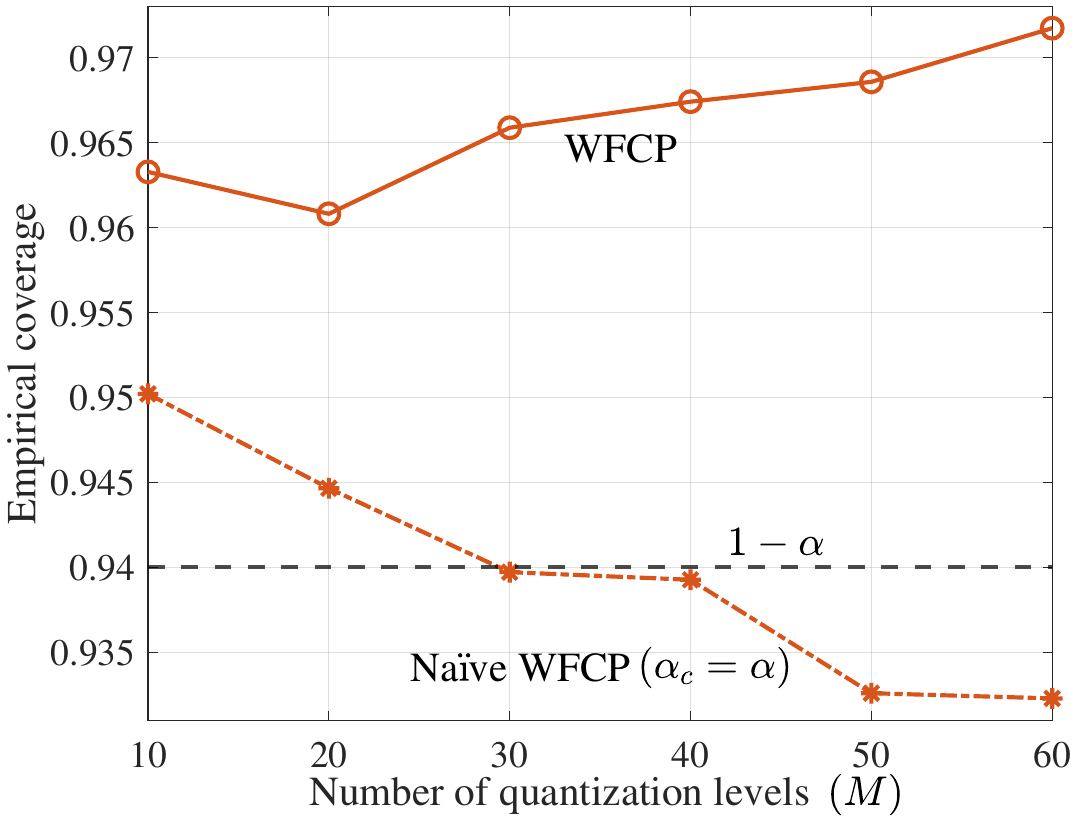}
	\includegraphics[width = 0.4\textwidth]{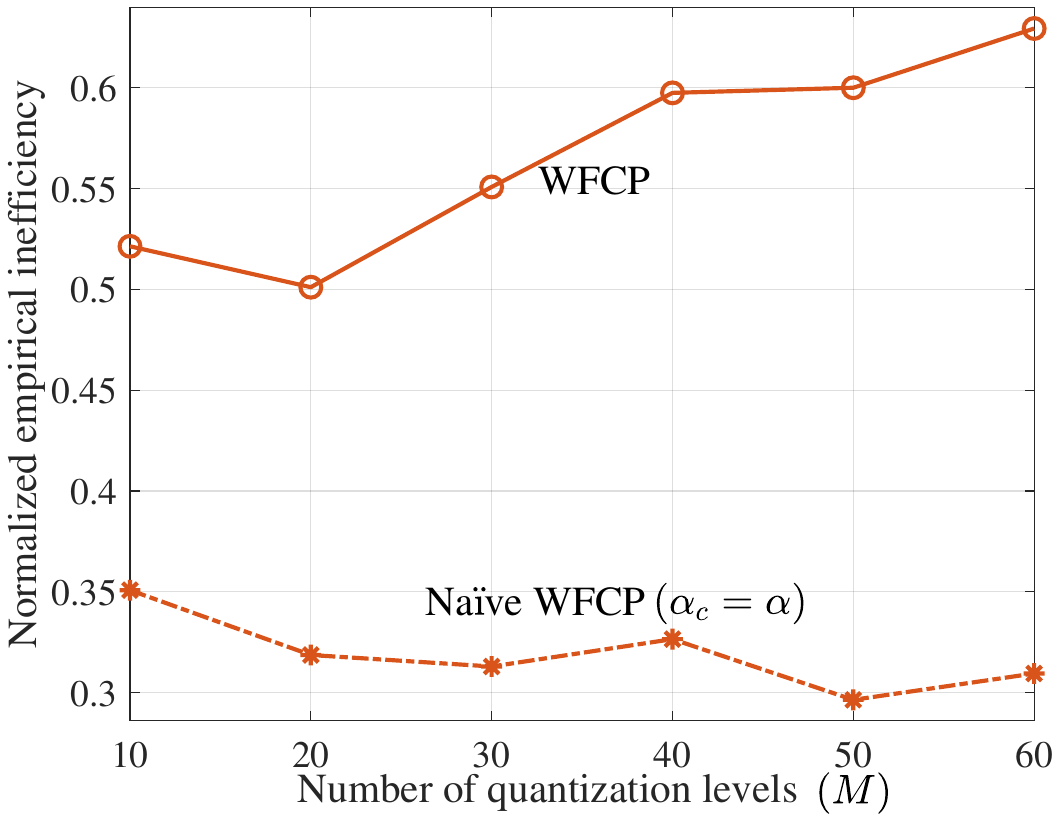}
	}
    \caption{{\color{blue}Empirical coverage and normalized empirical inefficiency of WFCP and na\"ive WFCP ($\alpha_c=\alpha$) versus the number $M$ of quantization levels with target unreliability level $\alpha=0.06$, $h^2_{\text{min}}=1$, number $T = 60$ of channel uses, number $K=30$ of devices, and $\text{SNR}=-10~\text{dB}$.}}
    \label{performance_M}
\end{figure*}
\begin{figure*}[htpb]
    \centering
    {
    \includegraphics[width = 0.4\textwidth]{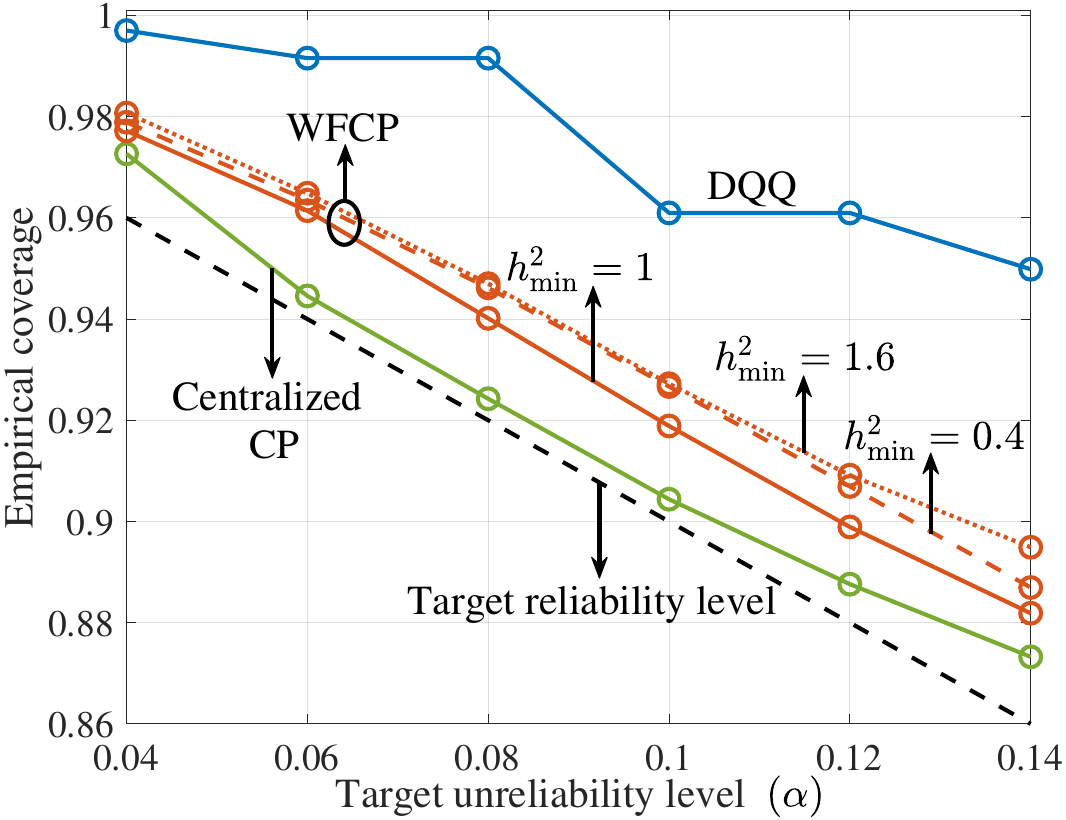}
    \includegraphics[width = 0.4\textwidth]{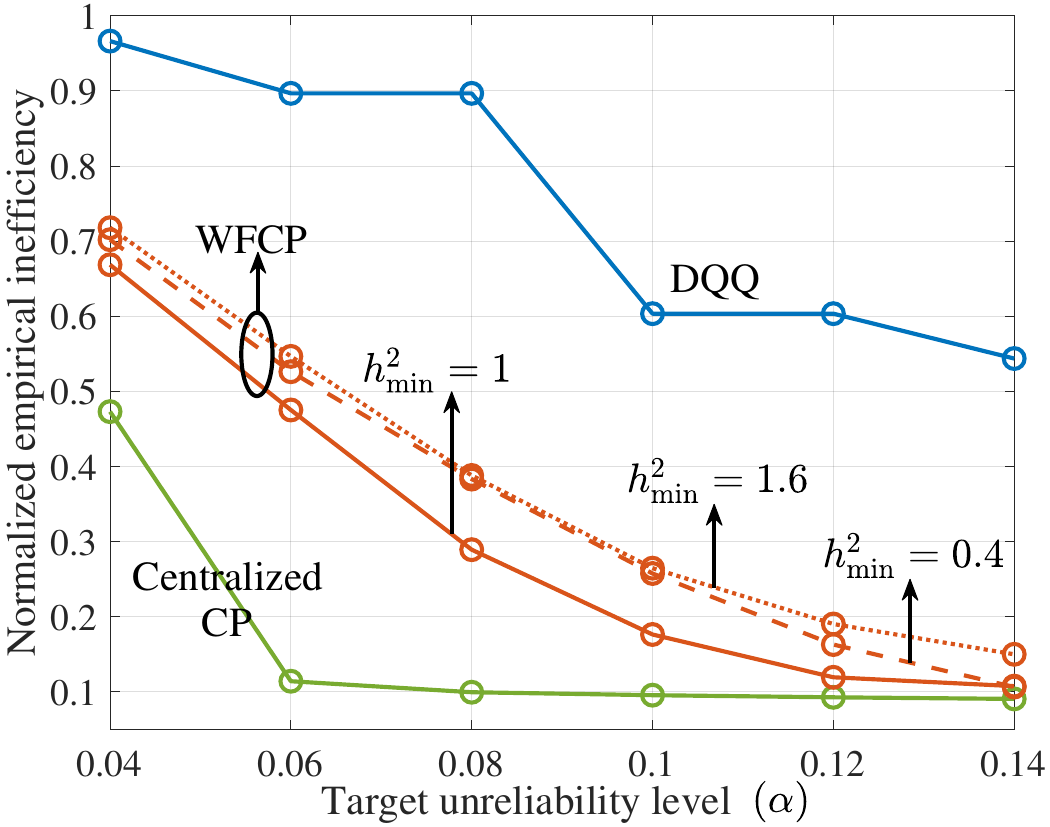}
	}
    \caption{{\color{blue}Empirical coverage and normalized empirical inefficiency of centralized CP, WFCP, and DQQ \cite{FedCP-QQ} versus the target unreliability rate $\alpha$ with $h^2_{\text{min}}=0.4, 1, 1.6$, number $M=20$ of quantization levels, number $T=60$ channel uses, number $K=20$ of devices, and $\text{SNR}=0~\text{dB}$.}}
    \label{performance_alpha}
\end{figure*}

\section{Experimental Settings and Results}\label{experiment}
In this section, we provide insights into the performance of the proposed WFCP via numerical results. We use as a benchmark the digital wireless implementation of FedCP-QQ \cite{FedCP-QQ}, abbreviated as DQQ, reviewed in Section \ref{Digital FedCP-QQ}.

\subsection{Setting}
{\color{blue}Following the experimental setting in reference \cite{angelopoulos2022private}, we use the CIFAR-10 data set, a standard benchmark for image classification involving $60000$ images classified using $C=10$ labels. We use $N^{\text{tr}}=50000$ data pairs for training the predictive model, while sampling $N=400$ points for calibration and another $N^{\text{te}}=400$ for testing from the remaining $10000$ data pairs. In the federated inference setup under study, each device holds $N/K$ calibration data points, while only the server has access to the $N^{\text{te}}$ test data points on which it wishes to generate reliable predictions.

The predictive model adopts the VGG-16 architecture \cite{simonyan2014very} with minor modifications. Specifically, the last layer is replaced by a linear layer with $C=10$ output neurons, followed by a softmax layer that outputs the conditional probability distribution $p(y|x)$.} We train the model using the standard federated gradient descent protocol \cite{mcmahan2017communication}. To this end, we divide the $N^{\text{tr}}$ training examples evenly across all devices. Following federated stochastic gradient descent, the server collects and averages the local gradients from a subset of the devices that are evaluated based on the respective local training data to update the model parameters via stochastic gradient descent. {\color{blue}Specifically, we utilize cross-entropy as the loss function, while adopting the SGD optimizer with a learning rate of $0.001$, momentum of $0.9$, and weight decay of $0.0005$, over $50$ epochs for the update procedure at the server. The final accuracy of the predictive model reaches $91.8\%$. As depicted in Fig. 1, the trained predictive model $p(y|x)$ is deployed at both devices and server.}

Since training is done offline and since our focus is on the inference phase, we do not account for constraints on the communication links during training. Training techniques that operate on noisy channels, as in \cite{liu2020privacy, zhu2019broadband, amiri2020machine, yang2020federated}, can be directly accommodated within the proposed federated inference framework.

{\color{blue}For the channel model, we consider Rayleigh fading channels, in which the channel powers are given as $h^2_k=0.5(a^2+b^2)$, with independent variables $a, b \sim \mathcal{N}(0,1)$.}

We adopt as performance measures the empirical coverage and empirical inefficiency, which are defined respectively as
\begin{align}\label{emp_cov}
    \text{Empirical coverage} = \frac{1}{N^{\text{te}}}\sum_{i=1}^{N^{\text{te}}}\mathds{1}\left(y_i \in \Gamma(x_i)\right)
\end{align}
and
\begin{align}\label{emp_inef}
    \text{Empirical inefficiency}= \frac{1}{N^{\text{te}}}\sum_{i=1}^{N^{\text{te}}}\left|\Gamma(x_i)\right|.
\end{align}

We run independent $400$ experiments to evaluate the above criteria, and obtain an average. {\color{blue}Each experiment involves sampling from the $10000$ data points not used for training to obtain $N$ calibration and $N^\text{te}$ test pairs.}

\begin{figure*}[t]
    \centering
    {
	\includegraphics[width = 0.4\textwidth]{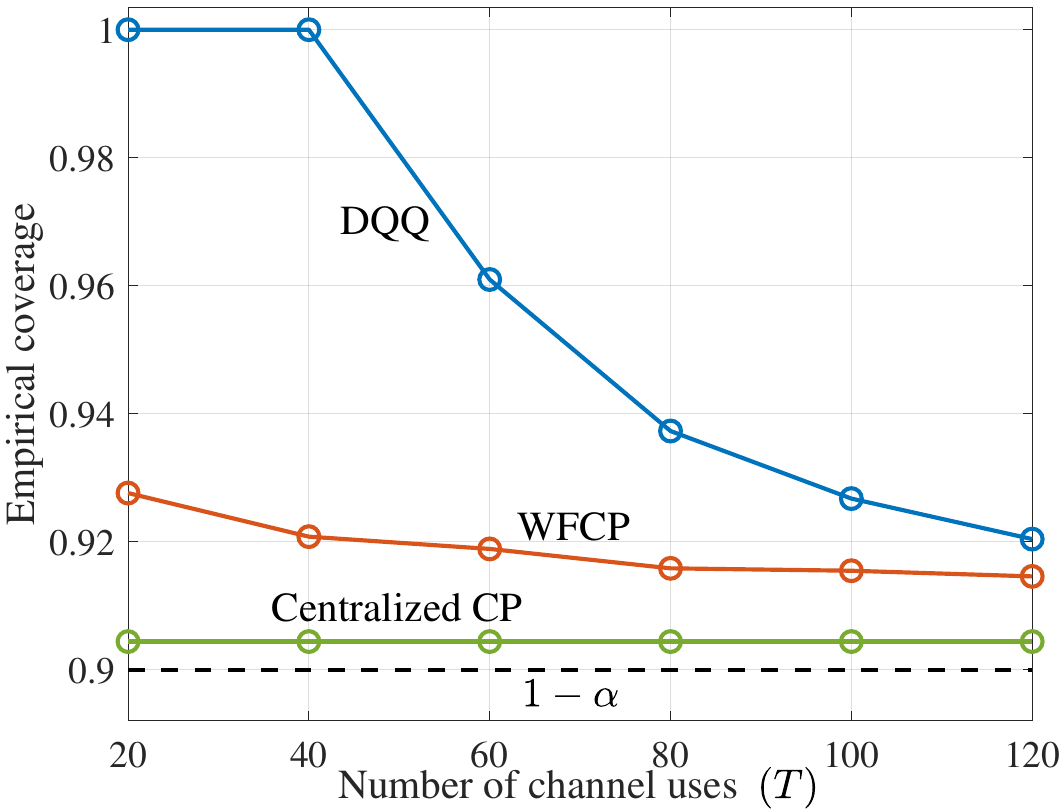}
	\includegraphics[width = 0.4\textwidth]{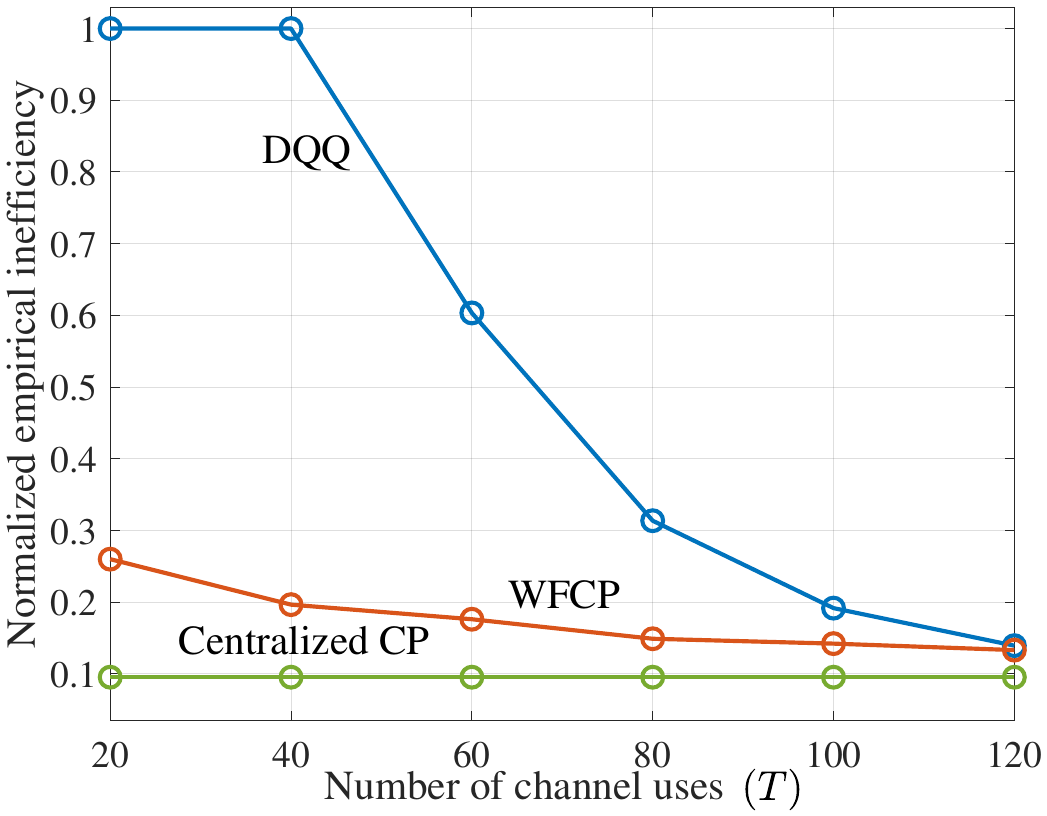}
	}
    \caption{{\color{blue}Empirical coverage and normalized empirical inefficiency of centralized CP, WFCP, and DQQ \cite{FedCP-QQ} versus the number $T$ of channel uses available with target unreliability level $\alpha=0.1$, $h^2_{\text{min}}=1$, number $M=20$ of quantization levels, number $K=20$ of devices, and $\text{SNR}=0~\text{dB}$.}}
    \label{performance_T}
\end{figure*}

\begin{figure*}[t]
    \centering
    {
	\includegraphics[width = 0.4\textwidth]{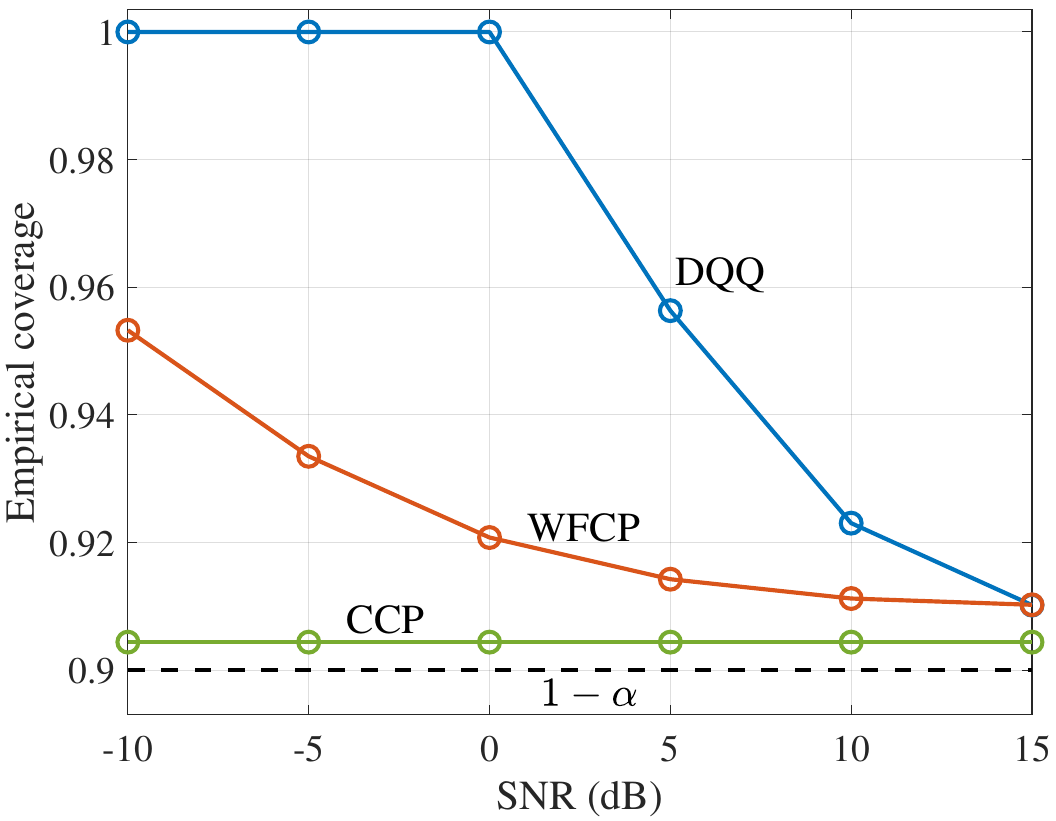}
	\includegraphics[width = 0.4\textwidth]{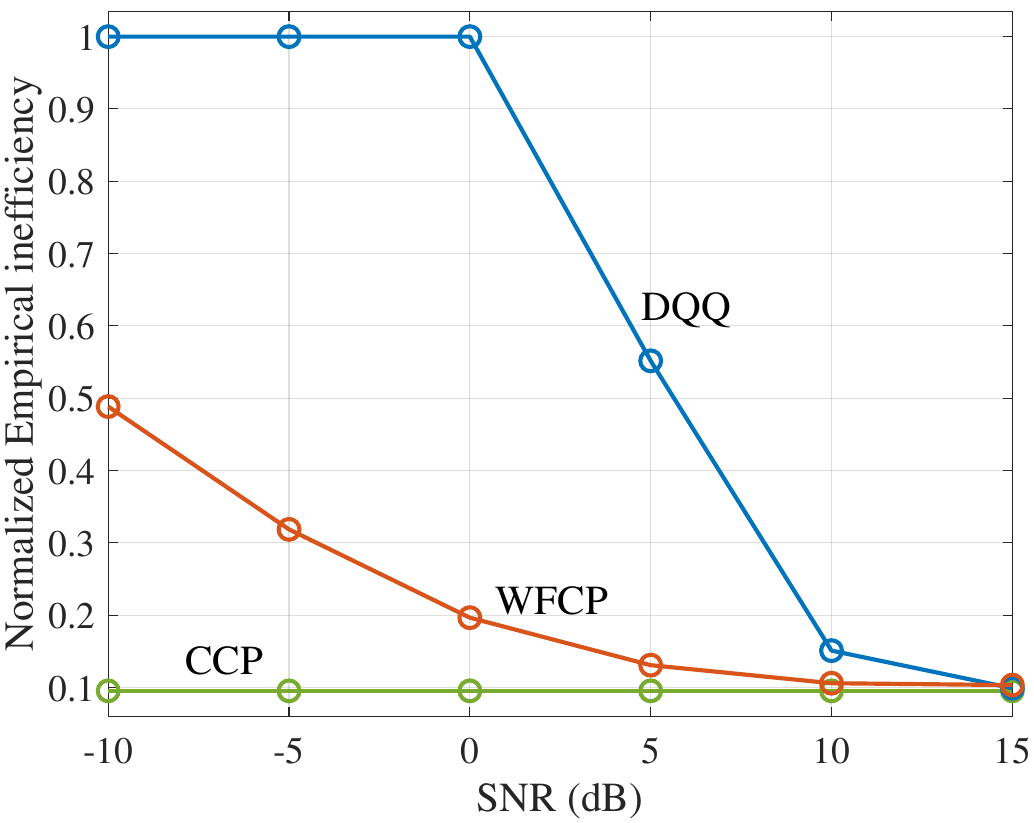}
	}
    \caption{{\color{blue}Empirical coverage and normalized empirical inefficiency of centralized CP, WFCP, and DQQ \cite{FedCP-QQ} versus SNR with target unreliability level $\alpha=0.1$, $h^2_{\text{min}}=1$, number $M=20$ of quantization levels, number $T=40$ of channel uses, and number $K=20$ of devices.}}
    \label{performance_SNR}
\end{figure*}

{\color{blue}
\subsection{On the Choice of the Number of Quantization Levels}
We start by focusing on the performance of the proposed WFCP scheme as a function of the number of quantization levels, $M$, for a fixed number $T=60$ of channel uses. This study is meant to substantiate the discussion in Sec. \ref{opt_M} on the optimal choice of $M$ as a trade-off between a less conservative correction, requiring a smaller $M$, and a larger resolution, calling for a larger $M$. For reference, we also consider a na\"ive implementation of WFCP which simply sets the target reliability level $1-\alpha_c$ in \eqref{noisy_quantile} to the true target $1-\alpha$ without considering the impact of channel noise.

Fig. \ref{performance_M} shows empirical coverage and empirical inefficiency for $\alpha=0.06$, $h^2_{\text{min}}=1$, $K=30$ devices, and $\text{SNR}=-10~\text{dB}$ as a function of $M$. As a first observation, confirming Theorem \ref{theorem_corr_quan}, WFCP achieves the target coverage reliability condition \eqref{coverage} for all quantization levels $M$. To obtain this goal, applying the corrected target reliability level $1-\alpha_c$ in \eqref{noisy_quantile} is essential. In fact, as also seen in the figure, the na\"ive implementation of WFCP fails to meet the coverage requirements \eqref{coverage} as soon as $M$ is sufficiently large, in which regime the performance is more sensitive to the presence of channel noise. For WFCP, the optimal value of $M$ in terms of inefficiency is observed to be around $M=20$, with smaller values causing a degraded performance due to an insufficient resolution and larger values generating an excessively conservative correction in \eqref{corr_term}.

\begin{figure*}[t]
    \centering
    {
	\includegraphics[width = 0.4\textwidth]{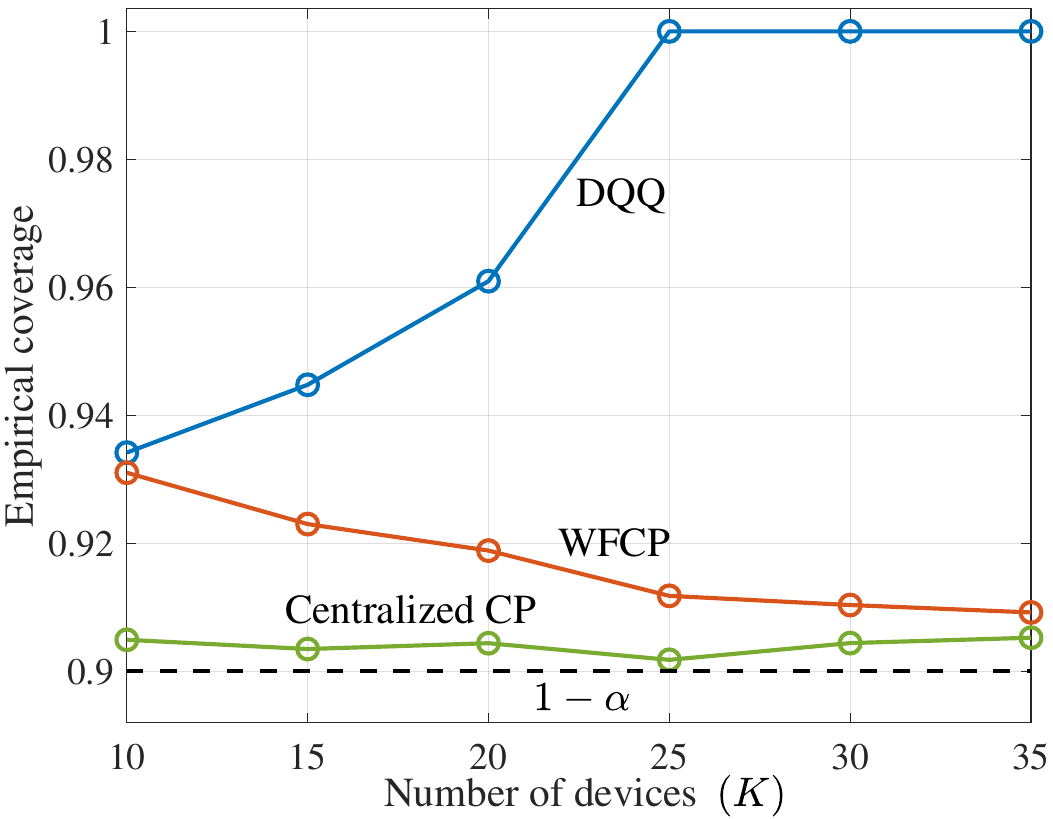}
	\includegraphics[width = 0.4\textwidth]{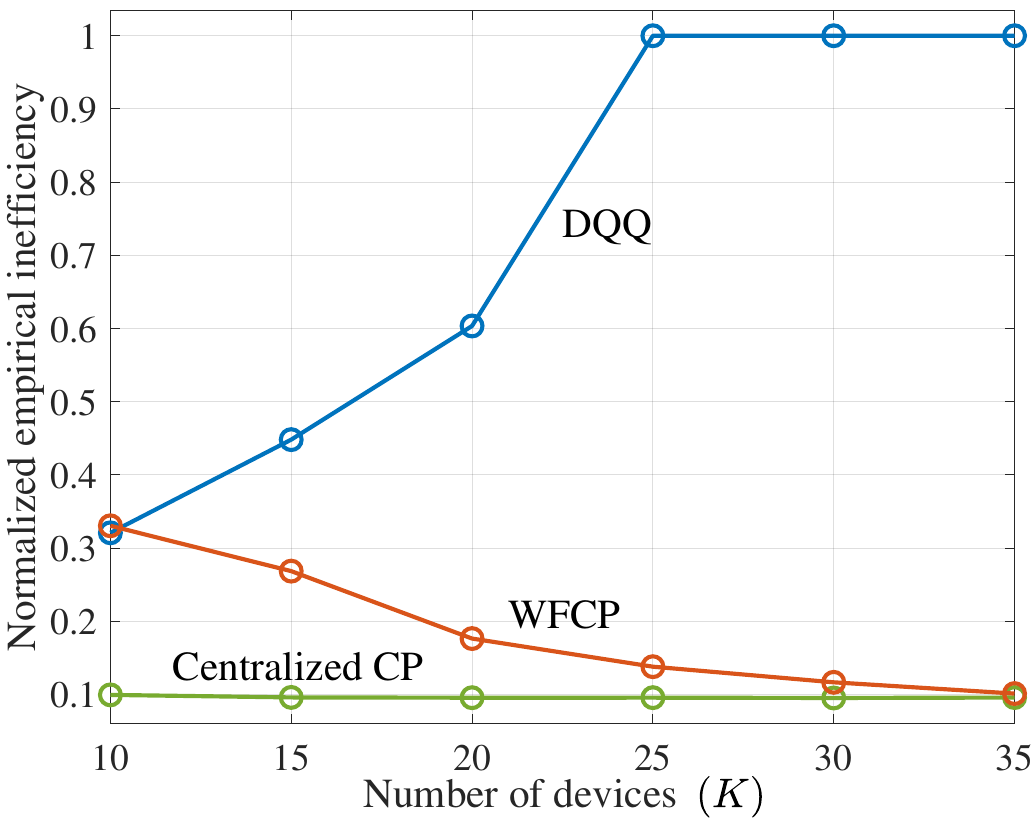}
    }
    \caption{\color{blue}{Empirical coverage and normalized empirical inefficiency of centralized CP, WFCP, and DQQ \cite{FedCP-QQ} versus the number $K$ of devices with $N_d=10$ per-device calibration data points, target unreliability level $\alpha=0.1$, $h^2_{\text{min}}=1$, number $M=20$ of quantization levels, number $T=60$ of channel uses, and $\text{SNR}=0~\text{dB}$.}}
    \label{performance_K_fixNd}
\end{figure*}

\subsection{Comparison between WFCP and DQQ}
We now turn to comparing the performance of WFCP and DQQ (Sec. \ref{Digital FedCP-QQ}). We start by evaluating empirical coverage and empirical inefficiency as a function of the target unreliability levels $\alpha$ in Fig. \ref{performance_alpha}. A larger unreliability level $\alpha$ corresponds to a smaller quantile threshold for DQQ in \eqref{DQQ_set_pre} and WFCP in \eqref{set_predictor}, yielding a smaller prediction set. WFCP consistently approximates the performance of the centralized CP, while DQQ yields prediction sets that are comparatively less informative. For example, for $\alpha=0.12$, the predicted set size of DQQ is nearly six times larger than the WFCP predicted set.

Furthermore, selecting the minimum channel power threshold $h_{\text{min}}^2$ entails a trade-off between the number of active devices, which decreases with $h^2_{\text{min}}$, and effective noise power \eqref{P_eff_noi}, which decreases with $h^2_{\text{min}}$. Compared to the threshold $h^2_{\text{min}}=1$, a lower threshold, here $h^2_{\text{min}}=0.4$, results in increased effective noise power $\sigma^2$ in \eqref{P_eff_noi}, while a higher threshold, here $h^2_{\text{min}}=1.6$, decreases the number $K_a$ of active devices as in \eqref{p_control_para}, with both choices yielding a less informative predicted set.}

We further evaluate the performance of WFCP and DQQ for different number $T$ of channel uses given a fixed number $M = 20$ of quantization levels. As seen in Fig. \ref{performance_T}, as $T$ increases, both methods maintain the target $(1-\alpha)$-coverage, while offering a decreasing inefficiency. This is because a larger $T$ weakens the effect of channel noise by reducing the probability of error $\epsilon$ in \eqref{P_error} for DQQ, and by improving the effective SNR in \eqref{eff_SNR} for WFCP. The proposed WFCP consistently outperforms DQQ, yielding highly informative prediction sets, with efficiency improvements being particularly evident in the regime of limited communication resources with low number $T$ of channel uses. As $T$ grows sufficiently large, the performance of both schemes approaches that of the centralized noiseless CP (Sec. \ref{background_CP}).

The performance gains of WFCP in the presence of limited communication resources are further explored in Fig. \ref{performance_SNR}, which evaluates the performance of WFCP and DQQ as a function of the SNR. As the SNR increases, the effective SNR in \eqref{eff_SNR} improves along with a decrease in the correction term in \eqref{corr_term}, resulting in a more informative predicted set, which approaches the performance of the centralized CP. In a similar manner, as the SNR improves, the probability of error $\epsilon$ in \eqref{P_error} for DQQ decreases, thereby generating a smaller-sized predicted set, which approaches the performance of WFCP for SNR levels around $15~\text{dB}$.

Fig. \ref{performance_K_fixNd} evaluates the performance of WFCP and DQQ when varying the number of devices, $K$.  Note that the number $N_d=10$ of per-device calibration data points is kept fixed, so that, as $K$ increases, the total number of calibration data points increases. For DQQ, as the number of devices increases, the inefficiency tends to increase. In fact, an increase in the number of devices leads to a higher error probability $\epsilon$ in \eqref{P_error}, which causes the average number of correctly received local quantiles, $K(1-\epsilon)$, to decrease.

In stark contrast, WFCP is observed to reduce the average predicted set size as the number $K$ of devices increases. Intuitively, this is due to the adoption of the TBMA protocol, which allows the on-air combination of signals transmitted by all the devices. At a technical level, this result is aligned with \eqref{corr_term}, which shows that the correction term is approximately independent of the number of calibration data per device and that it is inversely proportional to the square of the number of devices, $K$. Accordingly, as $K$ grows, the corrected target reliability level $1-\alpha_c$ approaches the true level $1-\alpha$, and the performance of WFCP approaches that of centralized CP.

\section{Conclusions and Outlooks}\label{conclusion}
This paper has introduced wireless federated conformal prediction (WFCP), the first protocol for the deployment of federated inference via CP in shared noisy communication channels. Like conventional centralized CP and some of the existing federated extensions of CP for noiseless channels, WFCP provably provides formal guarantees of reliability, indicating that the predicted set produced at the server contains the true output with any target probability. WFCP builds on type-based multiple access (TBMA), a communication protocol that allows the estimate of a global histogram from distributed observations with a bandwidth that scales with the resolution of the histogram and not with the number of devices. The key technical challenge tackled by this paper is the definition of a novel quantile correction approach that ensures the reliability of the set predictor despite the presence of channel noise. The theoretical analysis of WFCP's reliability performance also offers valuable insight into the choice of critical design parameters, such as the number of quantization levels.
Simulation results further substantiate the advantage of the proposed WFCP scheme over existing strategies, particularly under constraints of limited communication resources and/or large number of devices. All in all, the proposed WFCP protocol provides a promising framework for implementing federated CP in wireless communication scenarios, thereby establishing a robust foundation for future exploration in this domain.


{\color{blue}The proposed WFCP can be directly extended to the scenario with heterogeneous data distributions and different sizes of calibration data sets across the devices by leveraging the results in \cite{lu2023federated}. To this end, assume that each device $k$ stores $N_k$ calibration data points sampled from a local data distribution $p^*_k(x,y)$. With a minor modification described next, WFCP can still guarantee the reliability condition \eqref{cov_bound} when the probability is evaluated with respect to the mixture 
\begin{equation}
    p^*(x,y)=\sum_{k\in\mathcal{K}_a}\frac{N_k}{\sum_{k\in\mathcal{K}_a}N_k}p_k^*(x,y)
\end{equation}
of the local data distributions. This distribution naturally arises when the test point $(x,y)$ is sampled from the local distribution of a randomly sampled active device, with device $k$ chosen with probability proportional to the size $N_k$ of the local data set \cite{lu2023federated}.

To guarantee such a reliability condition, it is sufficient to choose the power control parameter $\gamma$ as in \eqref{gamma} with $\max_{k\in\mathcal{K}_a}N_k$ in lieu of $N_d$ and to select the threshold in \eqref{set_predictor} as the $\lceil(1-\alpha_c)(N_a+K_a)\rceil$-th smallest value, instead of the $\lceil(1-\alpha_c)(N_a+1)\rceil$-th smallest value, among the $N_a=\sum_{k\in\mathcal{K}_a}N_k$ NC scores. As noted in paper \cite{lu2023federated}, this scheme requires the condition $\lceil(1-\alpha_c)(N_a+K_a)\rceil \leq N_a$.

Another interesting direction for research is to devise a differentially private implementation of WFCP, potentially leveraging the idea of channel noise as a masking mechanism \cite{liu2020privacy}.}

\appendix
\subsection{Proof of Theorem 1}\label{theorem1}
{\color{blue}Given any fixed channel realization $\left\{h_k\right\}_{k=1}^K$ and a corresponding $h^2_{\text{min}}$, we assume that a subset $\mathcal{K}_a$ of devices are active. In this section, we denote the $N_a=K_aN_d$ quantized calibration NC scores from the $K_a$ active devices as $s_i=q(s(x_i,y_i))$ for $i=1, \ldots, N_a$ and quantized NC score for the \textit{true} test pair $(x,y)$ as $s_{N_a+1}=q(s(x,y))$.
We also introduce two sets of $N_a+1$ NC scores. The first includes both calibration and test NC scores, i.e.,
\begin{equation}\label{set1}
    \mathcal{S}^*=\{s_i\}^{N_a+1}_{i=1},
\end{equation}
while the second replaces the test NC score $s_{N_a+1}$ with the maximum NC score value $S_M$, i.e., 
\begin{equation}\label{set2}
    \mathcal{S}^{\text{max}}=\{s_i\}^{N_a}_{i=1}\cup\{S_M\}.
\end{equation}

For such a \textit{genie-aided} set $\mathcal{S}^*=\{s_1,\ldots,s_{N_a+1}\}$} that has access to the test NC score, we use the bag notation $\lbag\mathcal{S}^*\rbag$ to refer to a set of numerical values of the NC scores, which excludes the identity of the data point to which each NC score is assigned. Furthermore, we write as $\pi(\mathcal{S}^*)$ the indices of the data points assigned to each element in the bag $\lbag\mathcal{S}^*\rbag$. We use the same notation for $\mathcal{S^{\text{max}}}$. Based on these definitions, the set $\mathcal{S}^*$ is unambiguously identified by the bag $\lbag\mathcal{S}^*\rbag$ and by the assignment $\pi(\mathcal{S}^*)$.

Finally, given the bag $\lbag\mathcal{S}^*\rbag$, we introduce the $M\times 1$ vector as $\boldsymbol{p}(\lbag\mathcal{S}^*\rbag)$, in which each $m$-th entry represents the fraction of NC scores in  $\lbag\mathcal{S}^*\rbag$ equal to quantization level $S_m$. With this definition, the vector {\color{blue}$\tilde{\boldsymbol{p}}^+$} in \eqref{r} can be equivalently defined as
{\color{blue}\begin{equation}\label{pbag}
    \tilde{\boldsymbol{p}}^+=\boldsymbol{p}(\lbag\mathcal{S^\text{max}}\rbag),
\end{equation}}and hence vector $\boldsymbol{r}$ in \eqref{r} as 
\begin{equation}\label{vbag}
    \boldsymbol{r}=\boldsymbol{p}(\lbag\mathcal{S^\text{max}}\rbag)+\tilde{\boldsymbol{z}}.
\end{equation}

In \eqref{pbag} and \eqref{vbag}, we have used the fact that histograms do not depend on the ordering of the defined set. Recall that we are interested in finding a lower bound on the probability \eqref{cov_bound}, which can be expressed as the expectation
\begin{align}\label{int}
    &\Pr\left(y \in \Gamma_{\alpha_c}^{\text{\rm{WFCP}}}(x|\boldsymbol{r})\right)\nonumber\\
    &=\mathbb{E}_{\mathcal{S}^*,\tilde{\boldsymbol{z}}}\left[\mathds{1}(y \in \Gamma_{\alpha_c}^{\text{\rm{WFCP}}}(x|\boldsymbol{r}))\right]\nonumber\\
    &=\mathbb{E}_{\mathcal{S}^*,\tilde{\boldsymbol{z}}}\left[\mathds{1}(s_{N+1}\leq S_{m_{\alpha_c}(\boldsymbol{p}(\lbag\mathcal{S}^{\text{max}}\rbag)+\tilde{\boldsymbol{z}})})\right]\nonumber\\
    &\geq\mathbb{E}_{\mathcal{S}^*,\tilde{\boldsymbol{z}}}\left[\mathds{1}(s_{N+1}\leq S_{m_{\alpha_c}(\boldsymbol{p}(\lbag\mathcal{S}^*\rbag)+\tilde{\boldsymbol{z}})})\right]\nonumber\\
    &=\mathbb{E}_{\tilde{\boldsymbol{z}},\lbag \mathcal{S}^* \rbag}\mathbb{E}_{\pi(\mathcal{S}^*)|\tilde{\boldsymbol{z}},\lbag \mathcal{S}^* \rbag}\left[\mathds{1}(s_{N+1}\leq S_{m_{\alpha_c}(\boldsymbol{p}(\lbag\mathcal{S}^*\rbag)+\tilde{\boldsymbol{z}})})\right],
\end{align}
where in the second equality we have used \eqref{set_predictor} and \eqref{vbag}, while for the third inequality, we have followed a standard trick of CP (see, e.g., Lemma 1 of \cite{tibshirani2019conformal}). This states that replacing any single value with the maximum value will never decrease the respective empirical quantile value. In the last equality, we have used the law of iterated expectations, which allows us to apply the expectations in the sequence as explained next.

\subsection*{\textbf{First step: Bounding $\mathbb{E}_{\pi(\mathcal{S}^*)|\tilde{\boldsymbol{z}},\lbag \mathcal{S}^* \rbag}[\cdot]$}}

We begin by studying the inner expectation over the ordering $\pi(\mathcal{S}^*)$ after conditioning on the bag $\lbag \mathcal{S}^* \rbag$ and the noise vector $\tilde{\boldsymbol{z}}$. In the following, we write $\boldsymbol{p}^*=\boldsymbol{p}(\lbag \mathcal{S}^* \rbag)$ to simplify the notation.  Recall that $S_1,\ldots,S_M$ are the $M$ quantization levels. From exchangeability of the data, we have the equality (see, e.g., \cite{tibshirani2019conformal})
\begin{equation}\label{s_N1}
    \Pr\big[s_{N+1}=S_i|\tilde{\boldsymbol{z}}, \lbag \mathcal{S}^*\rbag \big]=p^*_i.
\end{equation}
It follows that we have the series of equalities
\begin{align}
    \mathbb{E}_{\pi(\mathcal{S}^*)|\tilde{\boldsymbol{z}},\lbag \mathcal{S}^* \rbag}&\left[\mathds{1}(s_{N+1}\leq S_{m_{\alpha_c}(\boldsymbol{p}^*+\tilde{\boldsymbol{z}})})\right]\nonumber\\
    =&\Pr\big[s_{N+1}\leq S_{m_{\alpha_c}(\boldsymbol{p}^*+\tilde{\boldsymbol{z}})}|\tilde{\boldsymbol{z}}, \lbag \mathcal{S}^*\rbag \big]\nonumber\\
    =&\hspace{-1em}\sum^{m_{\alpha_c}(\boldsymbol{p}^*+\tilde{\boldsymbol{z}})}_{i=1}\hspace{-1em}p^*_i\nonumber\\
    \geq& \min \left\{1,1-\alpha_c -\hspace{-1em}\sum^{m_{\alpha_c}(\boldsymbol{p}^*+\tilde{\boldsymbol{z}})}_{i=1}\hspace{-1em}\tilde{z}_i\right\},
\end{align}
where the inequality follows from the definition of $m_{\alpha_c}(\boldsymbol{p}(\lbag\mathcal{S}^{\text{max}}\rbag)+\tilde{\boldsymbol{z}})$ in \eqref{index_quantile} and the $\min\{\cdot\}$ operator is introduced to account for the case $m_{\alpha_c}(\boldsymbol{p}^*+\tilde{\boldsymbol{z}})=M$.

\subsection*{\textbf{Second step: Bounding ($\mathbb{E}_{\tilde{\boldsymbol{z}}|\lbag \mathcal{S}^* \rbag}[\cdot]$)}}
We now marginalize over the noise vector $\tilde{\boldsymbol{z}}$ given the bag $\lbag\mathcal{S}^*\rbag$. Given the bag $\lbag\mathcal{S}^*\rbag$ we have
\begin{align}\label{LB}
    \mathbb{E}&_{\tilde{\boldsymbol{z}}| \lbag\mathcal{S}^*\rbag}\mathbb{E}_{\pi(\mathcal{S}^*)|\tilde{\boldsymbol{z}},\lbag\mathcal{S}^*\rbag}\left[\mathds{1}\left(y \in \Gamma_{\alpha_c}^{\text{\rm{WFCP}}}\left(x|\boldsymbol{r}\right)\right)\right]\nonumber\\
    &\geq \mathbb{E}_{\tilde{\boldsymbol{z}}|\lbag\mathcal{S}^*\rbag}\left[\min \left\{1,1-\alpha_c -\hspace{-1em}\sum^{m_{\alpha_c}(\boldsymbol{p}^*+\tilde{\boldsymbol{z}})}_{i=1}\hspace{-1em}\tilde{z}_i\right\}\right]\nonumber \\
    &= 1+\frac{1}{2}\mathbb{E}_{\tilde{\boldsymbol{z}}|\lbag\mathcal{S}^*\rbag}\left[-\alpha_c - \hspace{-1em} \sum^{m_{\alpha_c}(\boldsymbol{p}^*+\tilde{\boldsymbol{z}})}_{i=1}\hspace{-1em}\tilde{z}_i-\Bigg|\alpha_c +\hspace{-1em} \sum^{m_{\alpha_c}(\boldsymbol{p}^* + \tilde{\boldsymbol{z}})}_{i=1}\hspace{-1em}\tilde{z}_i\Bigg|\right],
\end{align}
in which we have used the identity $\min\{x,y\}=y+\frac{x-y-|x-y|}{2}$ to obtain the last equality.

We now note that the index $m_{\alpha_c}(\boldsymbol{p}^*+\tilde{\boldsymbol{z}})$ depends on $\tilde{\boldsymbol{z}}$ and that, once conditioned on $\lbag\mathcal{S}^*\rbag$ it depends only on the realization of the sequence $\tilde{z}_1,\ldots,\tilde{z}_{m_{\alpha_c}(\boldsymbol{p}^*+\tilde{\boldsymbol{z}})}$. Therefore $m_{\alpha_c}(\boldsymbol{p}^*+\tilde{\boldsymbol{z}})$ is a stopping time for the sequence $\tilde{z}_1,\tilde{z}_2,\ldots$, and we are allowed to invoke first Wald's identity \cite{wald2004sequential} to obtain
\begin{equation}
    \mathbb{E}_{\tilde{\boldsymbol{z}}|\lbag\mathcal{S}^*\rbag}\left[\sum^{m_{\alpha_c}(\boldsymbol{p}^*+\tilde{\boldsymbol{z}})}_{i=1}\hspace{-1em}\tilde{z}_i\right]
    =\mathbb{E}_{\tilde{\boldsymbol{z}}|\lbag\mathcal{S}^*\rbag}\Big[m_{\alpha_c}(\boldsymbol{p}^*+\tilde{\boldsymbol{z}})\Big]\mathbb{E}[\tilde{z}_1]=0,
\end{equation}
where the last equality follows from the noise being zero mean. Furthermore, from Wald's second identity \cite{wald2004sequential}, we have 
\begin{align}
    \mathbb{E}_{\tilde{\boldsymbol{z}}|\lbag\mathcal{S}^*\rbag}&\left[\left(\sum^{m_{\alpha_c}(\boldsymbol{p}^*+\tilde{\boldsymbol{z}})}_{i=1}\hspace{-1em}\tilde{z}_i\right)^2\right]\nonumber\\
    
    \leq& \sigma^2 \mathbb{E}_{\tilde{\boldsymbol{z}}|\lbag\mathcal{S}^*\rbag}\left[m_{\alpha_c}(\boldsymbol{p}^*+\tilde{\boldsymbol{z}})\right]\leq \sigma^2M,
\end{align}
which will be used next.
Applying Jensen's inequality $\mathbb{E}[|x|]^2\leq \mathbb{E}[x^2]$, we can further bound \eqref{LB},
\begin{align}\label{tight LB}
    &1+\frac{1}{2}\mathbb{E}_{\tilde{\boldsymbol{z}}|\lbag\mathcal{S}^*\rbag}\left[-\alpha_c - \hspace{-1em} \sum^{m_{\alpha_c}(\boldsymbol{p}^*+\tilde{\boldsymbol{z}})}_{i=1}\hspace{-1em}\tilde{z}_i-\Bigg|\alpha_c +\hspace{-1em} \sum^{m_{\alpha_c}(\boldsymbol{p}^* + \tilde{\boldsymbol{z}})}_{i=1}\hspace{-1em}\tilde{z}_i\Bigg|\right] \nonumber\\
    & \geq 1 - \frac{\alpha_c}{2}-\frac{1}{2}\sqrt{\mathbb{E}_{\tilde{\boldsymbol{z}}|\lbag\mathcal{S}^*\rbag}\bigg[\alpha_c^2 + 2\alpha_c \hspace{-1em}\hspace{-0.5em}\sum^{m_{\alpha_c}(\boldsymbol{p}^* + \tilde{\boldsymbol{z}})}_{i=1}\hspace{-1.5em}\tilde{z}_i+ \bigg(\sum^{m_{\alpha_c}(\boldsymbol{p}^* + \tilde{\boldsymbol{z}})}_{i=1}\hspace{-1em}\tilde{z}_i\bigg)^2\bigg]} \nonumber\\
    &\geq 1 - \frac{\alpha_c}{2}-\frac{1}{2}\sqrt{\alpha_c^2 + \sigma^2M}.
\end{align}

Accordingly, we have
\begin{align}\label{final}
    \mathbb{E}_{\tilde{\boldsymbol{z}}| \lbag\mathcal{S}^*\rbag}&\mathbb{E}_{\pi(\mathcal{S}^*)|\tilde{\boldsymbol{z}},\lbag\mathcal{S}^*\rbag}\left[\mathds{1}\left(y \in \Gamma_{\alpha_c}^{\text{\rm{WFCP}}}\left(x|\boldsymbol{r}\right)\right)\right]\nonumber\\
    &\geq1 - \frac{\alpha_c}{2}-\frac{1}{2}\sqrt{\alpha_c^2 + \sigma^2M}.
\end{align}

\subsection*{\textbf{Final step: Bounding ($\mathbb{E}_{\lbag\mathcal{S}^* \rbag}[\cdot]$)}}
The final step follows directly from \eqref{final}, in which the lower bound does not depend on the bag $\lbag\mathcal{S}^*\rbag$. This gives that 
\begin{equation}
    \Pr\left(y \in \Gamma_{\alpha_c}^{\text{\rm{WFCP}}}(x|\boldsymbol{r})\right) \geq 1 - \frac{\alpha_c}{2}-\frac{1}{2}\sqrt{\alpha_c^2 + \sigma^2M}.
\end{equation}

Therefore, to satisfy the target coverage rate $1-\alpha$, we set the corrected unreliability level as
\begin{equation}\label{final2}
    \alpha_c = \alpha - \frac{\sigma^2 M}{4 \alpha}.
\end{equation}

\bibliographystyle{IEEEtran}
\bibliography{TBMA_FL_CP_revision.bib}

\end{document}